\documentstyle[prb, aps, epsf]{revtex}
\begin{document}
\title{
Valley current filtering and reversal by parallel side contacted armchair nanotubes
}
\author{Ryo Tamura}
\address{Faculty of Engineering, Shizuoka University, 3-5-1 Johoku, 
Hamamatsu 432-8561, Japan}

\maketitle

\begin{abstract} 
The intertube conductance $G$ 
of the parallel side contacted armchair nanotubes
is calculated by Landauer's formula 
as a function of the Fermi level $E$. 
When the intertube difference in the dope strength
is large enough,
a resonant peak dominant over the others 
appears in the $E$-$G$ curve.
The overlap length and the interlayer configuration do not influence the resonant energy.
The intervalley transmission
at the resonant peak
works as a reverser
and a filter of the valley current.

\end{abstract}

\twocolumn

\section{introduction}

Valleys -- the $K$ and $K'$ corner points in the two dimensional Brillouin zone --
are crucial for distinguishing properties
of the graphene (Gr) and
the monolayer transition metal dichalcogenide (TMD).\cite{valley-review}
An intervalley 
imbalance in the electron-hole recombination brings about the polarized light-emitting in the pn junction of TMD.\cite{TMD}
The valley current (VC) -- an intervalley difference in current --
does not necessarily accompany the charge current. \cite{pure-current} 
Owing to the high mobility, the Gr is more suitable for the VC technology
than the TMD.\cite{graphene-review} 
The Y-shaped bilayer Gr
with oppositely charged top and back gates\cite{top-back-gate}
separates the $K$ current from the $K'$ current.
The bubble structure\cite{bubble} and
line defects\cite{disclination-pair} in the single layer Gr
transmit only one of the $K$ and $K'$ currents.
We call these systems VC filters (VCFs) 
because we obtain 
the single valley current 
from simultaneous incidence of the $K$ and $K'$ currents.
On the other hand, the Gr superlattice structure (GrS) \cite{graphene-precession} and partially overlapped Gr (po-Gr) \cite{gra-gra-junction} work as the VC reversers (VCR) 
that cause large intervalley transmission rates.
Each of these systems, however,
can be only one of the VCF and the VCR,
while a structure that works both as the VCF
and as the VCR is suitable for the integration of the VCF with the VCR
in the device processing.

The carbon nanotube (NT) is similar to the Gr when the Gr satisfies the periodic boundary condition
along the NT circumference. \cite{NT-review} 
Since the interlayer current is much smaller than the intralayer current
in the multiwall NT and the multilayer Gr, 
one might consider that the interlayer current 
is irrelevant to the VCF and the VCR.
In the parallel side contacted nanotube (ps-NT)
and the telescoped NT (t-NT) schematically shown by Fig. 1,
however, the interlayer current is equal to the intralayer current.
\cite{side-tele,tamura-2010,tamura-2012,tamura-2019,slow-decay-side-contact-ref} In addition, the overlap length can be controlled
by mechanical motion of the attached piezo electrode. \cite{telescope-mechanical} 
When the NTs are armchair nanotubes (ANTs),
the valley indexes 
are translated into 
the symmetry index $\sigma=+,-$ 
concerning
the mirror plane on the ANT axis.
In this case, the VCR  producing the $\sigma$ VC
is realized when $T_{\sigma,-\sigma} \simeq 1$
 where $T_{\sigma,\sigma'}$ denotes the transmission rate from valley $\sigma'$
to valley $\sigma$.
Since $T_{+,+}+T_{-,+} \leq 1 $ 
and $T_{+,+}+T_{+,-} \leq 1 $,
a necessary condition for the VCR is (I) $T_{+,+} \simeq 0$.
It is noteworthy that condition (I) is compatible with
 the condition of VCF, $T_{\sigma,\sigma}+T_{\sigma,-\sigma} \simeq 1 \gg T_{-\sigma,\sigma}
+T_{-\sigma,-\sigma}$.

According to the result of the t-ANT,
condition (II) $|\varepsilon| \gg w_{+,+}$
is necessary for condition (I)
where
$\varepsilon$ is the {\it intertube} site energy difference 
of the tight binding model (TB)
and $w_{+,+}$ denotes the interlayer Hamiltonian element
corresponding to $T_{+,+}$.\cite{tamura-2010}
Reflecting the relatively large interlayer area
shown by the right panels of Fig. 1,
the t-ANT has a much larger $w_{+,+}$ than the ps-ANT.
In the range $|\varepsilon| \leq $ 0.5 eV,
the t-ANT cannot be the VCR because
the large $w_{+,+}$ conflicts with
condition (II).
Though the relatively small $w_{+,+}$ of the ps-ANT is favorable to achieving condition (II),
the discussion about the nonzero $\varepsilon$ is limited to the t-ANT.
In the experiment, the nonzero $\varepsilon$ can be induced by the encapsulated dopants, while it cannot be much larger than 0.5 eV.\cite{Iodine,Eu-wire,NearlyFree}  This paper discusses whether the ps-ANT works as the VCR and
 the VCF  with the realistic $\varepsilon$.
Evidence of the VC other than  nonlocal resistance ($R_{\rm NL}$)  
 \cite{non-local-resistence} is 
 also a target of this 
 paper as interpretation of $R_{\rm NL}$
  in terms of the VC is still controversial. \cite{VC-discussion}

This paper is organized as follows.
Section II defines the TB Hamiltonian.
Section III shows the notation of the transmission rate
for the discussion about the VCR and the VCF.
In Sec. IV, the approximate
formulas of the transmission rates are derived from the TB Hamiltonian.
Section V defines
the overlap integrals between the two ANTs
to clarify the relationship between the transmission rate
and the wave function
.
In Sec. VI, the origin of the VCR and the VCF
is elucidated by the approximate formulas
and the overlap integrals.
A summary and a conclusion are presented in 
Sec. VII, where we comment on the experimental fingerprint
of the VCR and the VCF.

\section{geometrical structure and TB Hamiltonian}
Figure 2 (a) 
schematically shows that
the ps-ANT is composed of
partially overlapped two ANTs.
We refer to the two ANTs as symbols $\downarrow$ and $\uparrow$.
Symbols L (R) denotes the region in tube $\downarrow$ ($\uparrow$) 
excluding the overlap region D.
Consider the honeycomb lattice $(\overline{y},\overline{z})$ 
on the $(y,z)$ plane where $ \overline{y}=\frac{\sqrt{3}a}{2} (m-\frac{(-1)^m}{6}-\frac{(-1)^j}{2})$, $ \overline{z}=\frac{a}{2}j$ with integers $j$, $m$ and the lattice constant $a=0.246$ nm.
Converting $\overline{y}$ into
the angles $\theta_\downarrow=\frac{\overline{y}}{r_\downarrow}$ and
$\theta_\uparrow=\frac{\overline{y}}{r_\uparrow} -\frac{2\pi}{3n_\uparrow}+\theta_{\rm d}$ with the tube radii $r_\xi=\frac{\sqrt{3}a}{2\pi}n_\xi$
and a small rotation $|\theta_{\rm d}| <\pi/n_\uparrow$,
we define the atomoc position $(x_\xi,y_\xi,z_\xi)$ of tube $\xi (=\downarrow,\uparrow)$ as 
$y_\xi=r_\xi\sin\theta_\xi$,
$x_\downarrow=r_\downarrow\cos\theta_\downarrow$, 
$x_\uparrow=r_0+r_\downarrow+r_\uparrow-r_\uparrow\cos\theta_\uparrow$
and $z_\downarrow=z_\uparrow-z_{\rm d}=\overline{z}$ 
with a small translation $|z_{\rm d}| < a/4$
and the interlayer distance $r_0= $ 0.31 nm.
In tube $\xi$, $1\leq m \leq 2n_\xi$.
Regions L, D and R correspond to the ranges $j \leq 0$,
$1\leq j \leq N-2$ and $N-1\leq j$, respectively,
where $N$ is an integer and the overlap length equals $\left(\frac{N}{2} -1\right)a-z_{\rm d}$.
Figure 2 (b) shows
the interlayer configuration
in case $n_\uparrow=2n_\downarrow$ and $ (\theta_{\rm d},z_{\rm d})=
(0,0)$.
The interlayer configuration is similar to the AB stacking when 
$z_{\rm d}=0$ and $\theta_{\rm d} =0,-2\pi/(3n_\uparrow)$. \cite{NT-NT-stacking}

According to the atomic position defined above,
we define
the TB Hamiltonian in the same way as Ref.\cite{Lambin}.
The TB equations
in region D are represented by
\begin{equation}
E\vec{c}_j^{\;(\rm D)}
=\sum_{\Delta j =-1}^1H^{(j,\Delta j)}
\vec{c}_{j+\Delta j}^{\;(\rm D)}
\label{TB-1}
\end{equation}
with the Hamiltonian matrix 
\begin{eqnarray}
H^{(j,\Delta j)} &=&
\left(
\begin{array}{cc}
h_{\downarrow}^{(j,\Delta j)}
, & 
0
\\

0, & h_{\uparrow}^{(j,\Delta j)}
\end{array}
\right)\nonumber \\
&& +
\left(
\begin{array}{cc}
0
, & 
W^{(j,\Delta j)}
\\

\;^tW^{(j+\Delta j,-\Delta j)}, & 0
\end{array}
\right).
\label{Hj} 
\end{eqnarray}
The first and second terms of Eq. (\ref{Hj}) are
intralayer $H_{\rm 0}^{(j,\Delta j)}$
and interlayer $V^{(j,\Delta j)}$ parts, respectively.
The nearest neighbor elements of $H_{\rm 0}$ and
the diagonal elements of $h_\uparrow^{(0,0)}$
equal $-t$ ($=-2.75$ eV) and $\varepsilon$, respectively,
while the other elements of $H_{\rm 0}$ are zero.

The nonzero value of $\varepsilon$ 
can be realized when the two NTs have different doping strengths.
For example, Fig. 1 schematically shows the pristine NT contacted with the iodine-encapsulated NT. \cite{Iodine}
The iodine atoms become negatively charged and exert the repulsive potential to the host NT wall electron,
while they have little influence on the other pristine NT.
When a single iodine atom is encapsulated,
the repulsive potential at a host NT's carbon site increases as the carbon site approaches the iodine atom.
On the other hand, when the iodine atoms are encapsulated densely and form the crystal, it is approximated by a positive constant $\varepsilon$ irrespective of the carbon site. 
We should also mention that the TB with the site energy shift is a standard tool for discussing
the doping effect.\cite{Farajian}

The element of $W^{(j,\Delta j)}$ 
becomes nonzero only when
the atomic distance $\overline{r}$ is shorter than
the cut-off distance $r_c=$ 0.39 nm.
The nonzero element is defined by 
$ t_1\exp[(r_1-\overline{r})/r_2]\cos(\theta^{\downarrow}+\theta^{\uparrow})$ 
with parameters $t_1=$ 0.36 eV, $r_1= $ 0.334 nm and $r_2=$ 0.045 nm.
The exact numerical calculation is performed in the
same way as Ref. \cite{tamura-2019}.
The same TB Hamiltonian is applied to both
the exact  and the approximate  calculations.

\newpage

\section{ notation of the transmission rate}
In Fig. 3 (a), (b), and (c), the rectangular area
represents the scattering region D between regions L and R.
The right and left going electrons
are represented by $\rightarrow$ and $\leftarrow$,
respectively.
The Bloch wave function of region $\mu (=$ L,R) with unit flow is denoted by
$\psi(\mu)$ with valley indexes $K,K'$
and the propagation direction indexes $\rightarrow, \leftarrow$.
The transmission rate from $\psi{\rm (L)}_{\flat,\rightarrow}$
to $\psi{\rm (R)}_{\sharp,\rightarrow}$
is denoted by $T_{\sharp, \flat}$
with valley indexes $\flat$ and $\sharp$.
The conductance equals $T_{K,K}+T_{K',K'}+T_{K',K}+T_{K,K'}$ 
in unit of the quantum conductance $2e^2/h$ according to Landauer's formula.
\cite{Datta} The wave functions of the isolated ANT
are classified into the symmetric (+)
and antisymmetric $(-)$ states with respect to
the mirror plane parallel to the ANT axis. 
Thick solid lines in Fig. 3 (d) illustrate one to one correspondence
between the symmetry indexes $+,-$ and the valley indexes $K, K'$ 
concerning the $\rightarrow$ waves.
Thus indexes $K, K'$ are replaced by $+,-$ in
the following sections.
In this section, however, we keep the notation $K, K'$ 
for comparison with other systems.

Figure 3 explains the transmission 
on condition (i) $T_{K',K}=1$.
In Fig. 3 (a), $\psi{\rm (L)}_{K,\rightarrow}$ 
is incident
and perfectly transmitted to $\psi{\rm (R)}_{K',\rightarrow}$.
Since the time reversal operation converts
$\psi{\rm (L)}_{K,\rightarrow}$ and $\psi{\rm (R)}_{K',\rightarrow}$
into
$\psi{(\rm L)}_{K',\leftarrow}$ and 
$\psi{(\rm R)}_{K,\leftarrow}$, respectively, 
it also converts Fig. 3 (a) into Fig. 3 (b).
In other words, the wave function of Fig. 3 (b)
is the complex conjugate of that of Fig. 3 (a).
It indicates that
$T_{\sharp, \flat}$ 
coincides with the
transmission rate from $\psi{\rm (R)}_{\sharp',\leftarrow}$
to $\psi{\rm (L)}_{\flat',\leftarrow}$ with the notation $(K')'= K$.
The VC of region $\mu (=$ L,R) is defined as $J(\mu)_v =J(\mu)_{ K}
-J(\mu)_{ K'}$ 
where $J(\mu)_\sharp$
denotes the flow
of electrons belonging to valley $\sharp (=K,K')$ of region $\mu$.
The VC is called 'pure' especially when it is unaccompanied by 
the charge current.
The pure VC can be generated and detected by the valley Hall effect
and the inverse valley Hall effect.\cite{VHE}
The VC of region $\mu$ becomes pure on condition 
$ J(\mu)_K +J(\mu)_{K'} = 0$.

Superposing Fig. 3 (a) on Fig. 3 (b),
we find Fig. 3 (c) where
$\psi{(\rm L)}_{K,\rightarrow}$
and
$\psi{(\rm R)}_{K,\leftarrow}$
are simultaneously
incident, and the positive pure VC of region L is converted 
into the negative pure VC of region R.
It manifests the VCR for the {\it positive} pure $J{\rm (L)}_v$.
In the same way, the ps-ANT works as the VCR for the {\it negative} pure 
$J{\rm (L)}_v$ on condition (ii) $T_{K,K'}=1$.
When conditions (i) $T_{K',K}=1$ and (iii) $T_{K,K'}=0$
are satisfied at the same time, on the other hand,
the ps-ANT works as the VCF producing the valley polarized
transmitted wave
$\psi{(\rm R)}_{K',\rightarrow}$ ($\psi{(\rm L)}_{K',\leftarrow}$)
from the valley unpolarized incidence
$\psi{(\rm L)}_{K,\rightarrow}+\psi{(\rm L)}_{K',\rightarrow}$
( $\psi{(\rm R)}_{K,\leftarrow}+\psi{(\rm R)}_{K',\leftarrow}$).
Interchanging $K$ and $K'$ in the discussion above, 
we can show the VCF producing
$\psi{(\rm R)}_{K,\rightarrow}$ and $\psi{(\rm L)}_{K,\leftarrow}$.

\section{Perturbative calculations }\label{sec-perturb}

The energies with a wave number $k$ 
of the periodic system of which the unit
cell is the same as region D 
are the eigen values
of the matrix $H(k)=H_{\rm 0}(k)+V(k) $ 
with the notation
\begin{equation}
\spadesuit(k)=
\left(
\begin{array}{cc}
\spadesuit^{(1,0)},
& 
\diamondsuit
\\
^t\diamondsuit^*,&
\spadesuit^{(2,0)} 
\end{array}
\right)
\label{spade-k} 
\end{equation}
where $\spadesuit= H, H_{\rm 0}, V$
and $\diamondsuit=\spadesuit^{(1,1)}+e^{-ika}\spadesuit^{(1,-1)}$.
The eigen values and eigen vectors of $H_{\rm 0}(k) $ 
are represented by
\begin{equation}
E_{\sigma,\xi}
=\sigma t \left( 2\cos\frac{ka}{2} -1 \right) 
+\varepsilon\delta_{\xi,\uparrow}
\label{disper-0}
\end{equation}
and
\begin{equation}
\vec{b}_{\sigma,\xi}^{\;[0](\zeta)}=
\left(
\begin{array}{c}
\vec{d}_{\sigma,\xi}^{\;[0]} \\
-\exp\left(i\frac{k}{2}a \right)\vec{d}_{\sigma,\xi}^{\;[0]}
\end{array}
\right),
\label{bd} 
\end{equation}
respectively, where $\delta_{\uparrow,\uparrow}=1,\delta_{\downarrow,\uparrow}=0$, 
\begin{equation}
\left(
\begin{array}{cc}
\vec{d}_{\sigma,\downarrow}^{[0]},& \vec{d}_{\sigma,\uparrow}^{[0]}
\end{array}
\right)
=\sqrt{2}
\left(
\begin{array}{cc}
\vec{g}_{\downarrow,\sigma}, & 0\\
0, & \vec{g}_{\uparrow,\sigma}
\end{array}
\right)
\label{dg}
\end{equation}
and
\begin{equation}
\;^t\vec{g}_{\xi,\sigma}=
\frac{1}{\sqrt{8n_\xi}}
(1,\sigma,1,\sigma,\cdots,1,\sigma)
\label{def-g} 
\end{equation}
with $\sigma$ being + $(-)$ for the symmetric
(antisymmetric) states.
The $m$'th component of the Eq. (\ref{def-g}) corresponds
to integer $m$ of the first paragraph of Sec. II.
The index $\zeta =\pm$ in Eq. (\ref{bd}) indicates
that $k \simeq \zeta 2\pi/(3a)$.
As is shown in Fig. 3 (d), $\zeta =\sigma$ ($\zeta =-\sigma$) 
for the $\leftarrow$ states ($\rightarrow$ states).
We introduce the factor $(-1)$ in the lower part of Eq. (\ref{bd})
in order to limit the range of $k$ to the Brillouin zone $|k| < \pi/a$.
The wave number $k$ satisfying Eq. (\ref{disper-0}) 
for the $\rightarrow$ states ($dE/dk > 0$) 
is approximated by
\begin{equation}
k_{\sigma,\xi} = \frac{ 2}{a}
\left(\frac{E-\varepsilon\delta_{\xi,\uparrow}}{\sqrt{3}t}-\sigma\frac{\pi}{3}\right)
\label{k-E} 
\end{equation}
when $k$ is close to $\pm 2\pi/(3a)$.

As we consider $V(k)$ to be the perturbation,
Eqs. (\ref{disper-0}) and (\ref{bd}) represent the zeroth order 
states. 
Equation (\ref{def-g}) is the same notation as Ref. \cite{tamura-2019}. The factor $\sqrt{2}$ in Eq. (\ref{dg}) is necessary
for the orthonormality,
$\;^t\!\left(\vec{b}_{\sigma,\xi}^{\;[0]}
\right)^*\vec{b}_{\sigma',\xi'}^{\;[0]}=\delta_{\sigma,\sigma'}\delta_{\xi,\xi'}
$. 
In contrast to the case of Ref. \cite{tamura-2019} ,
the linear bands $k_{\sigma,\uparrow}$ and $k_{\sigma,\downarrow}$ 
are not degenerate and the first order shift of the energy 
$^t(\vec{b}_{\sigma,\xi}^{[0]})^{*}V(k)\vec{b}_{\sigma,\xi}^{[0]}$
equals zero.
The first order shift $\vec{b}^{[1]}$ in the wave function 
is represented by

\renewcommand{\arraystretch}{2}
\begin{equation}
\left(
\begin{array}{c}
^t \vec{b}_{+,\downarrow}^{[1],(\zeta)}
\\
^t \vec{b}_{-,\downarrow}^{[1],(\zeta)}
\end{array}
\right)
=
2
\left(
\begin{array}{cc}
\frac{-w_{+,+}^{(-\zeta)}}{\varepsilon}
, & \frac{w_{-,+}^{(-\zeta)}}{2E-\varepsilon} \\
\frac{w_{+,-}^{(-\zeta)}}{2E-\varepsilon}
, & \frac{-w_{-,-}^{(-\zeta)}}{\varepsilon}
\end{array}
\right)
\left(
\begin{array}{c}
^t \vec{b}_{+,\uparrow}^{[0],(\zeta)}
\\
^t \vec{b}_{-,\uparrow}^{[0],(\zeta)}
\end{array}
\right)
\label{K1}
\end{equation}
\renewcommand{\arraystretch}{1}
and
\renewcommand{\arraystretch}{2}
\begin{equation}
\left(
\begin{array}{c}
^t \vec{b}_{+,\uparrow}^{[1],(\zeta)}
\\
^t \vec{b}_{-,\uparrow}^{[1],(\zeta)}
\end{array}
\right)
=
2
\left(
\begin{array}{cc}
\frac{w_{+,+}^{(\zeta)}}{\varepsilon}
, & \frac{w_{+,-}^{(\zeta)}}{2E-\varepsilon} \\
\frac{w_{-,+}^{(\zeta)}}{2E-\varepsilon}
, & \frac{w_{-,-}^{(\zeta)}}{\varepsilon}
\end{array}
\right)
\left(
\begin{array}{c}
^t \vec{b}_{+,\downarrow}^{[0],(\zeta)}
\\
^t \vec{b}_{-,\downarrow}^{[0],(\zeta)}
\end{array}
\right)
\label{K2} 
\end{equation}
\renewcommand{\arraystretch}{1}
where 
\begin{equation}
w_{\sigma',\sigma}^{(-)}
\equiv \frac{1}{2}\; ^t(\vec{b}_{\sigma',\uparrow}^{[0],(+)})^{*}V
\left ( \frac{2\pi}{3a} \right)\vec{b}_{\sigma,\downarrow}^{[0],(+)}
\label{def-w2} 
\end{equation}
and $w_{\sigma',\sigma}^{(+)}=w_{\sigma',\sigma}^{(-)*}$
in the same way as Ref. \cite{tamura-2019}. The explicit definition is 
\begin{equation}
w_{\sigma',\sigma}^{(\zeta)} =
\sum_{j=1}^2\sum_{m=1}^{2n_\downarrow}
\sum_{m'=1}^{2n_\uparrow}
\frac{ (W^{(j,0)}-\widetilde{W}^{(j,1),(\zeta)})_{m,m'}}{8\sigma^{m+1}(\sigma')^{m'+1}
\sqrt{n_\uparrow n_\downarrow}}
\label{def-w} 
\end{equation}
where integers $m,m'$ correspond to $m$ of the first paragraph of
Sec. II and
\begin{equation}
\widetilde{W}^{(j,1)(\pm )}=
e^{\pm i \frac{\pi}{3}}W^{(j,1)}
+
e^{ \mp i \frac{\pi}{3}}W^{(j,-1)}.
\label{wideW}
\end{equation}

Figure 4 is a schematic diagram of Eq. (\ref{k-E}) in case $k \simeq 2\pi/(3a)$
and helps us to derive Eqs. (\ref{K1}) and (\ref{K2}).
The dashed horizontal line indicates the Fermi level $E$ .
The approximate eigenvectors $\vec{b}^{[0],(+)}+\vec{b}^{[1],(+)}$ up to the first order are calculated along the corresponding dotted vertical lines that indicate the Fermi wave numbers 
$-k_{+,\xi}$ and $k_{-,\xi}$
.
The four closed symbols on
the dashed horizontal line correspond to 
the unperturbed wave function $\vec{b}^{[0],(+)}$ 
at the energy $E$
while the first order shift $\vec{b}^{[1],(+)}$ 
is the linear combination of the states indicated by the open symbols.
For example, the closed diamond
corresponds to $\vec{b}^{[0],(+)}_{-,\downarrow}$ 
while the first order shift $\vec{b}^{[1],(+)}_{-,\downarrow}$ 
caused by $V(k_{-,\downarrow})$ is the superposition of the open diamonds.
\cite{JJ-Sakurai}
Since $ k_{-,\downarrow} \simeq 2\pi/(3a)$, 
the matrix elements of $V(k_{-,\downarrow})$ are approximated by those of
$V(2\pi/(3a))$.

When we neglect modes other than the $\vec{b}^{[0]}+\vec{b}^{[1]}$ 
states defined
by Eqs. (\ref{bd}), (\ref{K1}) and (\ref{K2}),
the solution of Eq. (\ref{TB-1}) 
can be approximated by the repetition of the
reduced vector $\vec{c}_j^{ \;\prime (\rm D)}$ as 
$(\vec{c}_j^{\;\prime (\rm D)},\vec{c}_j^{\;\prime (\rm D)},
\cdots, \vec{c}_j^{ \;\prime (\rm D)})$.
The neglect of the evanescent modes is justified when 
the overlap length is much larger than
the NT radii, i.e., $N \gg n_\downarrow,n_\uparrow$
because the decay length is 
about the NT radius at most.
The reduced vector $\vec{c}_j^{\;\prime (\rm D)}$ is represented
by
\begin{equation}
\vec{c}_{j}^{\;\prime
(\rm D)}
=
\left(
U_{\rm D}^{[0]}+U_{\rm D}^{[1]}
\right)
\sum_{s=\pm} 
\Xi^{s(j+1)}
\vec{\gamma}_s^{(\rm D)}
\label{cj-D} 
\end{equation}

The sign $s=+$ ($s=-$) 
corresponds to the $\rightarrow$ states
($\leftarrow$ states).
The matrix $U_{\rm D}^{[n]}$ 
denotes the $n$'th order term for $\vec{c}_{-1}^{\;\prime ({\rm D})}$, 
\begin{equation}
U_{\rm D}^{[n]}
=
\left( 
\vec{d}^{\;\prime[n],(-)}_{+,\downarrow},
\;
\vec{d}^{\;\prime[n],(+)}_{-,\downarrow},
\;
\vec{d}^{\;\prime [n],(-)}_{+,\uparrow},
\;
\vec{d}^{\;\prime [n],(+)}_{-,\uparrow} \right)
\label{def-UD}
\end{equation}
where $\vec{d}^{\;\prime[0],(\zeta)}_{\sigma,\xi}$
is defined
by Eq. (\ref{dg}) of which $\vec{g}_{\xi,\sigma}$ is replaced by $
\vec{g}^{\;\prime}_{\xi,\sigma}=(1/\sqrt{8n_\xi})\;^t(1,\sigma)$.
Replacing $\vec{b}_{\sigma,\xi}^{[n],(\zeta)}$ 
with $\vec{d}_{\sigma,\xi}^{\;\prime [n],(\zeta)}$ in Eqs. (\ref{K1}) and (\ref{K2}), we also obtain $\vec{d}^{\;\prime [1],(\zeta)}_{\sigma,\xi}$.
Accordingly
the index $\zeta$ in Eq. (\ref{def-UD}) is necessary only when $n=1$
as $\vec{d}^{\;\prime[0],(+)}= \vec{d}^{\;\prime[0],(-)}$ and $\vec{d}^{\;\prime[1],(+)}= \vec{d}^{\;\prime[1],(-)*}$.
The phase factor matrix
is derived from 
Eq. (\ref{k-E}) as
\begin{equation}
\Xi
\equiv
\left(\begin{array}{cc}
\beta_1 \Omega
, & 0\\
0, & \beta_2 \Omega
\end{array}
\right)
\label{def-Xi} 
\end{equation}
where $\beta_1=e^{i\frac{E}{\sqrt{3}t}},
\beta_2 =e^{i\frac{E-\varepsilon}{\sqrt{3}t}}$ and
\begin{equation}
\Omega
=\left(
\begin{array}{cc}
e^{i\frac{2}{3}\pi},& 0 \\
0,& e^{-i\frac{2}{3}\pi}
\end{array}
\right)\;\;.
\label{def-Omega} 
\end{equation}
Replacing $U^{[1]}_{\rm D}$ with zero in
Eq. (\ref{cj-D}), we obtain

\begin{equation}
\vec{c}_j^{\;\prime
(\rm L)}
=
\frac{v_0}{\sqrt{n_\downarrow}}\\
\sum_{s=\pm} 
\left[ \beta_1
\Omega\right]^{s(j+1)}
\vec{\gamma}_s^{(\rm L)}
\label{cj-L}
\end{equation}
for region L $(j \leq -1$),
\begin{equation}
\vec{c}_j^{\;\prime
(\rm R)}
=
\frac{v_0}{\sqrt{n_\uparrow}}\\
\sum_{s=\pm} 
\left[\beta_2
\Omega\right]^{s(j-N+1)}
\vec{\gamma}_s^{(\rm R)}
\label{cj-R}
\end{equation}
for region R $(j \geq N-1)$, where
\begin{equation}
v_0
=\frac{1}{2}\left(
\begin{array}{cc}
1,& 1 \\
1,& -1
\end{array}
\right).
\end{equation}

Equations (\ref{cj-D}), (\ref{cj-L}) and (\ref{cj-R}) enable us to transform
the boundary conditions 
\begin{eqnarray}
\left(
\begin{array}{cc}
\vec{c}^{\;\prime(\rm D,\downarrow)}_{0}
,\;&
\vec{c}^{\;\prime
(\rm D,\uparrow)}_{N-2}
\\
\vec{c}^{\;\prime (\rm D,\downarrow)}_{-1}
,\;&
\vec{c}^{\;\prime
(\rm D,\uparrow)}_{N-1}
\\
\vec{c}^{\;\prime (\rm D,\uparrow)}_{-1},\;&
\vec{c}^{\;\prime
(\rm D,\downarrow)}_{N-1}
\end{array}
\right)
=\left(
\begin{array}{cc}
\vec{c}^{\;\prime(\rm L)}_{0} ,\;&
\vec{c}^{\;\prime (\rm R)}_{N-2}
\\
\vec{c}^{\;\prime (\rm L)}_{-1}
,\;& 
\vec{c}^{\;\prime(\rm R)}_{N-1}
\\
0,\;& 0
\end{array}
\right)
\label{boundary} 
\end{eqnarray}
into formulas
$(^t\!\vec{\gamma}^{\rm (D)}_+,\;^t\!\vec{\gamma}^{\rm (L)}_-)^t\!X_{\rm L}
=
-(^t\!\vec{\gamma}^{\rm (D)}_-,\;^t\!\vec{\gamma}^{\rm (L)}_+)^t\!X_{\rm L}^*
$
and
$(^t\!\vec{\gamma}^{\rm (D)}_-\Xi^{-N},\;^t\!\vec{\gamma}^{\rm (R)}_+)^t\!X_{\rm R}
=
-(^t\!\vec{\gamma}^{\rm (D)}_+\Xi^{N},\;^t\!\vec{\gamma}^{\rm (R)}_-)^t\!X_{\rm R}^*
$
.
Partitioning $U_{\rm D}^{[n]}$ as $^tU_{\rm D}^{[n]}=
(^tU_{\rm L}^{[n]}/\sqrt{n_\downarrow},\;^tU_{\rm R}^{[n]}/\sqrt{n_\uparrow})
$ , we derive the $n$'th order of $X_\mu$ matrix as
\begin{equation}
X_{\mu}^{[n]}
=\left(
\begin{array}{cc}
U_\mu^{[n]}\Xi_0, & -v_0\Omega^*\delta_{n,0}\\
U_\mu^{[n]}, & -v_0\delta_{n,0}\\
U_{-\mu}^{[n]}, & 0\\
\end{array}
\right)
\label{def-X} 
\end{equation}
with the notation $-$L =R and $-$R =L.
In the derivation of Eq. (\ref{def-X}),
$E/t$ and $\varepsilon/t$ are neglected
and $\Xi$ is replaced by
\begin{equation}
\Xi_0
=\left(
\begin{array}{cc}
\Omega,& 0 \\
0, & \Omega
\end{array}
\right).
\label{Xi0} 
\end{equation}
The results of the perturbative calculation
on the $S$ matrix $S_\mu=-X_\mu^{-1}X_\mu^*$ 
are 
\begin{eqnarray}
S_{\rm L}
&=& 
\left(
\begin{array}{ccc}
0
,&-F_+^*,
& {\bf 1}_2
\\
-\;^tF_+^*,& -{\bf 1}_2, & -\;^tF_+\\
{\bf 1}_2,& -F_+
& 0\\
\end{array}
\right)
\label{SL}
\end{eqnarray}

and
\begin{eqnarray}
S_{\rm R}
&=& 
\left(
\begin{array}{ccc}
-{\bf 1}_2
,&-F_-,
& -F_-^*
\\
-\;^tF_-,& 0, & {\bf 1}_2\\
-\;^tF_-^*,& {\bf 1}_2,
& 0\\
\end{array}
\right)
\label{SR}
\end{eqnarray}
where 
\renewcommand{\arraystretch}{2}
\begin{equation}
F_\pm
=
2
\left(
\begin{array}{cc}
\frac{\mp w_{+,+}^{(\pm)}}{\varepsilon}
, & \frac{w_{-,+}^{(+)}}{2E-\varepsilon} \\
\frac{w_{+,-}^{(-)}}{2E-\varepsilon}
, & \frac{\mp w_{-,-}^{(\mp)}}{\varepsilon}
\end{array}
\right)\;\;.
\label{F+-}
\end{equation}
\renewcommand{\arraystretch}{1}
Here ${\bf 1}_n$
denotes the unit matrix of $n$ dimension
and the derivation of Eqs. (\ref{SL}) and (\ref{SR}) is
shown in Appendix A.
Equations (\ref{SL}) and (\ref{SR})
satisfy the time reversal symmetry
$\;^tS_\mu =S_\mu$,
while the unitarity
$S_\mu^*S_\mu ={\bf 1}_6$ holds
up to the first order.
Partitioning $S_\mu$ matrixes in Eqs. (\ref{SL}) and (\ref{SR}) 
as
\begin{equation}
S_\mu=
\left(
\begin{array}{cc}
r_\mu
, &
^tt_\mu \\
t_\mu, &
0
\\
\end{array}
\right),
\label{smu}
\end{equation}
The transmission rate block
in the $S$ matrix of the double junction
L-D-R corresponding to incidence from region L 
is represented by
\begin{equation}
t_{\rm RL}= t_{\rm R} 
\Xi^N
\sum_{m=0}^{\infty}\;
\left(r_{\rm L}\Xi^N r_{\rm R}
\Xi^N\right)
^m\;^t t_{\rm L}\;. 
\label{tRL} 
\end{equation}
In the formula up to the first order,
$t_{\rm RL} =(\beta_1\beta_2\Omega)^Nt'_{\rm RL}\Omega^N$ where
\begin{eqnarray}
t'_{\rm RL} &= &
\;^tF^*_+(\beta_1\Omega)^N-
\;^tF_+(\beta_1\Omega)^{-N} \nonumber \\
& &
+
(\beta_2\Omega)^N\;^tF_- -(\beta_2\Omega)^{-N}\;^tF_-^*
\label{first-order}
\end{eqnarray}
As the first (second) row and column of Eq. (\ref{first-order}) correspond to
symmetric $+$ (antisymmetric $-$) channel,
the transmission rate $|(t_{\rm RL})_{\sigma',\sigma}|^2$ from $(\sigma,$L)
to $(\sigma',$R) is represented by
$T_{\sigma',\sigma}= 16 \frac{|w_{\sigma',\sigma}|^2}{3t^2} Q_{\sigma',\sigma}$
where
\begin{equation}
Q_{\sigma',\sigma}=\left(\frac{\sin(\eta_{\sigma',\sigma}^{(-\sigma\sigma')}N)}{\eta_{\sigma',\sigma}^{(-\sigma\sigma')}}
\cos(\eta^{(\sigma\sigma')}_{\sigma',\sigma}N-\phi_{\sigma',\sigma})\right)^2
\label{perturb-T-total}
\end{equation}
with the beat wave numbers $ \eta_{\sigma',\sigma}^{(\pm)} \equiv (k_{\sigma,\downarrow}\pm
k_{\sigma',\uparrow})a/4$
and the phase $\phi_{\sigma',\sigma}$ of the parameter 
$w_{\sigma',\sigma}^{(\sigma)}=|w_{\sigma',\sigma}^{(\sigma)}|\exp(i\phi_{\sigma',\sigma})$. 
The explicit formulas are
\begin{eqnarray}
T_{\sigma,\sigma}&=& \frac{64 |w_{\sigma,\sigma}|^2}{\varepsilon^2}
\sin^2\left(\frac{\varepsilon N}{2\sqrt{3}t}\right) \nonumber \\
& & \times \cos^2\left[
\left(\frac{\widetilde{E}}{\sqrt{3}t}-\frac{\pi}{3}\sigma\right)N-
\phi_{\sigma,\sigma}\right]
\label{perturb-T-diag}
\end{eqnarray}
for the intravalley transmission ($\sigma\sigma'=+$) 
and
\begin{eqnarray}
T_{-\sigma,\sigma}
&=&\frac{16 |w_{-\sigma,\sigma}|^2}{\widetilde{E}^2}
\sin^2\left(\frac{\widetilde{E} N}{\sqrt{3}t}\right) \nonumber \\
& & \times \cos^2\left[\left( \frac{\varepsilon}{2\sqrt{3}t}-\frac{\pi}{3}\sigma\right)N
-\phi_{-\sigma,\sigma}
\right]
\label{perturb-T-offdiag}
\end{eqnarray}
for the intervalley transmission ($\sigma\sigma'=-$) 
with 
$\widetilde{E} \equiv E-\frac{\varepsilon}{2}$ being the energy measured from 
the cross point 
between the $E^{[0]}_\downarrow$ and
$E^{[0]}_\uparrow$ bands.
Equations (\ref{def-w}) and (\ref{wideW}) show that
$|w_{\sigma',\sigma}|$ is an even function of $z_{\rm d}$.
Accordingly
$\left. \frac{\partial |w_{-\sigma,\sigma}|}{\partial z_{\rm d}} \right|_{z_{\rm d}=0}=0$
and
\begin{eqnarray}
\left. \frac{\partial T_{-\sigma,\sigma}}{\partial z_{\rm d}} 
\right|_{z_{\rm d}=0}
&=&
\left.
\left( \frac{\partial \phi_{-\sigma,\sigma}}{\partial z_{\rm d}} 
\frac{\partial T_{-\sigma,\sigma}}{\partial \phi_{-\sigma,\sigma}} 
\right)
\right|_{z_{\rm d}=0}
\nonumber \\
& =& 
\left. \frac{\partial \phi_{-\sigma,\sigma}}{\partial z_{\rm d}} 
\right|_{z_{\rm d}=0}C_{\sigma}
\sin\left(
\frac{\varepsilon N}{\sqrt{3}t}-\frac{2}{3}\pi\sigma N
\right) 
\label{tuika}
\end{eqnarray}
where $C_\sigma=16\frac{|w_{-\sigma,\sigma}|^2}{\widetilde{E}^2}\sin^2\left(
\frac{\widetilde{E}N}{\sqrt{3}t}
\right) $. We will refer to Eq. (\ref{tuika}) in discussion about the dependence of $G$ on $z_{\rm d}$.

Equation (\ref{perturb-T-total})
reproduces the analytical formulas of Ref. \cite{tamura-2019}
only when $\sigma\sigma'=-1$.
An interpolation 
between Eq. (\ref{perturb-T-diag}) and the formula of Ref. \cite{tamura-2019}
is
\begin{equation}
T_{\sigma,\sigma}=\frac{Y_\sigma}{Y_\sigma+\left[1-
\frac{16 |w_{\sigma,\sigma}|^2}{\Gamma_\sigma^2}\sin^2\left(\frac{\Gamma_\sigma N}{2\sqrt{3}t}\right) \right]^2}
\label{SL2} 
\end{equation} 
where
\begin{eqnarray}
Y_{\sigma}&=& \frac{64 |w_{\sigma,\sigma}|^2}{\Gamma_\sigma^2}\sin^2\left(\frac{\Gamma_\sigma N}{2\sqrt{3}t}\right)
\cos^2\left(
\eta_{\sigma,\sigma}^{(+)}N- \phi_{\sigma,\sigma}
\right)\;,
\label{Ysigma} 
\end{eqnarray}
and 
\begin{equation}
\Gamma_\sigma=\sqrt{\varepsilon^2+16|w_{\sigma,\sigma}|^2}\;\;.
\label{gamma} 
\end{equation}
Equations (\ref{perturb-T-diag}) and (\ref{Ysigma})
are transformed into each other
by the replacement $\varepsilon \leftrightarrow \Gamma_\sigma$.
As $|w_{\sigma,\sigma}/\varepsilon|$ decreases,
Eq. (\ref{SL2}) approaches Eq. (\ref{perturb-T-diag}).
On the contrary, 
Eq. (\ref{SL2}) 
coincides with the formula of Ref. \cite{tamura-2019} 
when $\varepsilon=0$.
In contrast to Eq. (\ref{perturb-T-diag}), Eq. (\ref{SL2}) is always lower than unity as the transmission rate should be.
We can derive Eq. (\ref{SL2}) in the absence of the intervalley parameters
$w_{-\sigma,\sigma}$ as is shown by Appendix B.

As is shown in Fig. 2 (a), the geometrical overlap length
equals $(N-2)$ in the unit of $a/2$ while 
the overlap length in the analytical formulas 
(\ref{perturb-T-diag}), (\ref{perturb-T-offdiag})
and (\ref{SL2})
seems $N$.
The difference between the two kinds of the overlap lengths
comes from the two margins at $j=-1, N-1$.
For the relation between the overlap length and the integer $N$, 
we should refer to 
the TB model of $(N-1)$ atoms aligned at
$z=ja/2$ ($j=0,1,2,\cdots,N-2)$
with a constant transfer integral $-t$.
The eigen value $E$ and $j$'th component $c_j$ 
of the eigen vector $\vec{c}$ of the TB secular equation 
are represented by $E=-2\gamma\cos (k_ma/2)$ and
$c_j=\sin(k_ma(j+1)/2)$, respectively, where $k_m= 2\pi m/(Na)$
with the integers $m$ under the boundary conditions $c_{-1}=0$ and $c_{N-1}=0$.
Here the system length in the formula of $k_m$ is $Na/2$ 
and less than the geometrical length $(N-2)a/2$.

\section{overlap integral analysis}
To search Eqs. (\ref{perturb-T-offdiag}) and (\ref{Ysigma}) for the characteristics of the wave function,
the sc-ANT is mimicked by the double ladder (DL) in Fig. \ref{Fig-ladder}.
The Hamiltonian of the DL is defined by Eq. (\ref{Hj})
where $n_\downarrow=n_\uparrow=1$ 
and all the elements of $W$ are zero except that $W^{(1,0)}_{2,1} =W^{(2,0)}_{2,1} =t_\bot$.
The intraladder transfer $-t$ and interladder transfer $t_{\bot}$
are indicated by the single line
and the double line, respectively. 
The sites with the interladder transfer $t_{\bot}$ 
are depicted as double circles.
Though the twisted DL in Fig. \ref{Fig-ladder} (a)
is topologically
identical to the DL with staggered $t_\bot$ in Fig. \ref{Fig-ladder} (b),
the essential period $a/2$ is clarified in the former 
while the latter is in harmony with the real geometric structure.
The secular equation $(E-H(k))\vec{b}=0$ of the DL is represented by
\begin{equation}
\left(
\begin{array}{cccc}
E, & -p, & 0, & 0 \\
-p, & E, & -t_{\bot}, & 0 \\
0, & -t_\bot, & E', & -p \\
0, & 0, & -p, & E'
\end{array}
\right)
\vec{d}_{\rm D}
=0
\label{overlap-secular} 
\end{equation}
where $p=2t\cos(ka/2)-t$ and $E'=E-\varepsilon$.
The eigen vector is represented by
\begin{equation}
\;^t\vec{d}_{\rm D}
=\left (1,\; q_{\rm r},\; q_\uparrow \left(\frac{E'}{E}q_{\rm r},\; 1 \right) \right)
\label{overlap-vecD} 
\end{equation}
where 
\begin{equation} (q_{\rm r}, q_\uparrow)=\left(\frac{E}{p},\; \frac{E^2-p^2}{t_\bot E'} \right)
\label{overlap-def-AB}
\end{equation}
\begin{equation}
p=\tau'\sqrt{\widetilde{E}^2+\frac{\varepsilon^2}{4}+\tau\sqrt{(\varepsilon^2+t_\bot^2)\widetilde{E}^2-\frac{\varepsilon^2t_\bot^2}{4}}}
\label{overlap-p} 
\end{equation}
$\tau=\pm 1$ and $\tau'=\pm 1$.
The relation between (\ref{overlap-vecD}) and Fig. \ref{Fig-ladder} 
is displayed in Fig. \ref{Fig-vecD}.
The squared overlap integral between the unperturbed $\xi \; (=\downarrow,\uparrow)$ region and
perturbed D region is calculated as
\begin{equation}
f_{\sigma,\xi} =
\frac{
\left|\;^t\vec{d}_{\rm D}\vec{d}^{\;\prime [0]*}_{\sigma,\xi}\right|^2
}
{
\left|\vec{d}_{\rm D}\right|^2\left|\vec{d}^{\;\prime[0]}_{\sigma,\xi}\right|^2
}
\label{overlap-f} 
\end{equation}
We speculate that the transmission rate $T_{\sigma',\sigma}$ reflects
the product of the squared overlap integrals defined by
\begin{equation}
I_{\sigma',\sigma}=2\sum_{\tau'=\pm}\sum_{\tau=\pm}f_{\sigma',\uparrow}f_{\sigma,\downarrow}
\label{def-I} 
\end{equation}
The necessary conditions for the transmission rate
$T_{\sigma',\sigma} \leq 1$ and $T_{\sigma,\sigma}+T_{-\sigma,\sigma} \leq 1$ 
are shared by $I$ as proved in Appendix C.

\section{Results and discussions}
The system $n_\downarrow=5, n_\uparrow =10$ is considered as
a typical example.
The effect of $n_\xi$ will
be discussed in the end of this section.
Firstly we discuss the AB stacking configuration $(\theta_{\rm d},z_{\rm d})=(0,0)$. 
When $z_{\rm d} =0$, Eq. (\ref{def-w}) is real and thus $\phi_{\sigma',\sigma}=0, \pi$.
The effect of nonzero $\theta_{\rm d}$, $z_{\rm d}$ will be shown latter.
Figure 7 shows $T_{-,-}$ 
as a function of $N$
for the energy $E=0.15$ eV.
It offers an archetypal example of the analysis of the $N$-$T_{\sigma,\sigma}$ curve by Eq. (\ref{SL2}).
The solid and dashed lines display
the exact results and approximate formula (\ref{SL2}), respectively,
for five values of $\varepsilon=0, 0.07,\cdots,0.28$ eV.
Table I shows $w_{-,-}$.
The residue $l$ of $N$ is defined by $l=N-3m$ 
with the integer $m$ under the condition $|l| \leq 1$. 
Since $|\eta_{\sigma',\sigma}^{(\sigma\sigma')}| 
\simeq \frac{\pi}{3}$ 
in Eq. (\ref{perturb-T-total}), the beat wave number $\eta_{\sigma',\sigma}^{(\sigma\sigma')}$ 
causes the rapid oscillation with the period $\Delta N =3$ .
In order to make the slow oscillation of the other 
beat wave number $\eta_{\sigma',\sigma}^{(-\sigma\sigma')}$ noticeable,
we fix the residue $l$ to 0, 1
and $-1$ in the upper, middle and lower panels, respectively,
in Fig. 7.
Overall agreement between the dashed and solid lines is fairly good.
For the quantitative comparison, 
the nodes of the dashed lines are listed in Table II.
The zero points of Eq. (\ref{SL2}) are represented by
$N=(6m-3-2l)\lambda_E/6$ and $N=m\lambda_\Gamma$
with $\lambda_E=\pi\sqrt{3}t/\widetilde{E}$, $\lambda_\Gamma=2\pi\sqrt{3}t/\Gamma$ and integers $m$.
Owing to the agreement between the solid and dashed lines,
the nodes of the solid lines in Fig. 7 can be also identified as those in Table II.
When $\varepsilon = 0.14$ eV, all the lines approach zero
at $N=200$ corresponding to the node $\lambda_\Gamma=201$.
When $\varepsilon=0.14 $ eV and $l=-1$, the nodes $\lambda_\Gamma=201$ 
and $5\lambda_E/6=156$ are close to each other, thus $T_{-,-}$ is
remarkably suppressed between the nodes $156 < N < 201$
in the bottommost panel.
The topmost panel also shows the similar suppression between the nodes
$\lambda_\Gamma=139$ and $\lambda_E/2=166$ when $\varepsilon=$ 0.21 eV.
We cannot discern the nodes 
$N=16.6, 21.7$ and 31.2 in the middle panel,
whereas they cause the apparent depletion of $T_{-,-}$ 
in the region $N<30$ compared to the other panels.
The approximate formula (\ref{SL2}) 
reproduces the suppression of the exact $T_{-,-}$ by the increase of $\varepsilon$.
The suppressed $T_{-,-}$ of $\varepsilon=0.21, 0.28$ eV
in the middle panel
is magnified in the inset.
The lines with circles correspond to case $\varepsilon =0.21$ eV.
Though Eq. (\ref{SL2}) overestimates (underestimates) $T_{-,-}$ when
$\varepsilon= 0.21$ eV and $N>50$ ($\varepsilon=0.28$ eV and $N>100$), 
the dashed and solid lines share the important characteristic that
they are remarkably smaller than those for $\varepsilon=0.14, 0.07, 0$ eV.
In the dashed lines of the inset, 
we can identify the nodes $\lambda_\Gamma=139, 105$, $\frac{\lambda_E}{6}=55.4$.
Although Eq. (\ref{SL2}) is verified when $\varepsilon =0$ or $|\varepsilon| \gg 4|w_{\sigma,\sigma}|$, agreement between the solid and dashed lines 
with $\varepsilon= 0.07$ eV manifests the effectiveness of
Eq. (\ref{SL2}) in the interpolated 
region $|\varepsilon| \simeq 4|w_{\sigma,\sigma}|$.
The same result is found in $T_{+,+}$ (not shown in Figures).

Equation (\ref{SL2}) is effective
not only in the ps-ANTs but also in the t-ANTs. 
In this paragraph, we concentrate our discussion into the 
intravalley transmission rate $T_{\sigma,\sigma}$ of the t-ANTs.
In Ref. \cite{tamura-2010}
, Author considered the averaged transmission rate $\overline{T}(N)=
\frac{1}{3}(T(N-2)+T(N)+T(N+2))$ in order to remove the rapid oscillations and derived 
\begin{equation}
\overline{T}_{\sigma,\sigma}{\rm (Ref. \;12)}
=16\frac{|w_{\sigma,\sigma}|^2}{\Gamma^2_\sigma}
\sin^2\left(\frac{\Gamma_\sigma N}{2\sqrt{3}t}\right)
\label{ref-2010}
\end{equation}
When $|\varepsilon| \gg 8|w_{\sigma,\sigma}|$, on the other hand, 
$Y_\sigma \ll 1$ and Eq. (\ref{SL2})
is close to $Y_\sigma$.
In that case, the averaged transmission rate $\overline{Y}_{\sigma}$ 
with the approximation $\cos^2(\eta^{(+)}_{\sigma,\sigma} N-\phi_{\sigma,\sigma}) \simeq
\cos^2(\pi N/3)$ 
is twice as large as Eq. (\ref{ref-2010}).
In this way, the multiple reflection ignored in Ref. \cite{tamura-2010} doubles $\overline{T}_{\sigma,\sigma}$
when $\overline{T}_{\sigma,\sigma} \ll 1$.
In the results displayed in Ref. \cite{tamura-2010} , we are aware that
Eq. (\ref{ref-2010}) is
only {\it half} of the exact averaged transmission rate when
$T_{\sigma,\sigma} \ll 1$.
The factor two in the formula $\overline{Y}_{\sigma}
\simeq \overline{T}_{\sigma,\sigma}{\rm (Ref.\;12})\times 2
$ has corrected this disagreement.\cite{footnote}

In order to satisfy the condition $T_{-\sigma,\sigma} \gg T_{\sigma,\sigma}$, $|\varepsilon|$ 
has to be large enough to suppress $T_{\sigma,\sigma}$.
In the following, we set $\varepsilon$ at 0.3 eV.
It is comparable to the experimentally evaluated value 0.6 eV.\cite{Iodine} 
Under the conditions
$|\varepsilon|, |\widetilde{E}| \ll \sqrt{3}t$ and $z_{\rm d} =0$, 
Eqs. (\ref{perturb-T-diag}) and (\ref{perturb-T-offdiag})
approximately satisfy the relations
\begin{equation}
T_{\sigma',\sigma}(\widetilde{E}, N)
=
T_{\sigma',\sigma}(-\widetilde{E}, N_E)
\label{relation-E} 
\end{equation}
and
\begin{equation}
T_{\sigma',\sigma}(\widetilde{E}, N)
|w_{-\sigma',-\sigma}|^2
=
T_{-\sigma',-\sigma}(\widetilde{E}, N_w)
|w_{\sigma',\sigma}|^2
\label{relation} 
\end{equation}
where $N_E$ and $N_w$ are related to $N=3m+l$
as $(N_E,N_w)=(3m-\sigma'\sigma l, 3m-l)$ 
with integers $m$ and residues $l=0,\pm 1$.
The origin of Eq. (\ref{relation-E}) can be traced back to
the beat wave number $\eta^{(+)}_{\sigma',\sigma}$.
Equation (\ref{SL2}) also
satisfies Eq. (\ref{relation-E}) under the same conditions,
while increase of $|w_{\sigma,\sigma}| $ weakens effectiveness of Eq. (\ref{relation}) for Eq. (\ref{SL2}).
In the following, however, we discuss the results for $\varepsilon =0.3$ eV
being much larger than $4|w_{\sigma,\sigma}|$. 
In that case, difference between Eqs. (\ref{SL2}) and (\ref{perturb-T-diag}) 
is so small that Eq. (\ref{SL2}) also approximately satisfies
Eq. (\ref{relation}).

Equations (\ref{perturb-T-offdiag}) and (\ref{SL2})
in case $l=1$ are exhibited in Fig. 8 with the range $ N=4,7,10,\cdots, 100$ and $ |E| < 0.3 $ eV.
The intravalley (\ref{SL2}) reaches a local maximum
when $N \simeq \lambda_\Gamma (m -\frac{1}{2})$ and
$E=\frac{\varepsilon}{2}+\pi\sqrt{3}t\left(m'+\frac{\sigma}{3}l\right)/N$
with integers $m$ and $m'$.
As for the local maximums i $\sim$ iv in Fig. 8, 
$m'=0,-1$ and $N \simeq \lambda_\Gamma/2$. On the other hand, 
v and vi denote the local maximums 
of the intervalley formula (\ref{perturb-T-offdiag}) 
that appear when $E=\varepsilon/2$ and 
$x\tan(x-\frac{\pi}{3}\sigma l) =1$ with the notation $x=\varepsilon N/(2\sqrt{3}t)$.
Figure 9 displays the exact numerical data corresponding to Fig. 8.
We are satisfied by the agreement between Fig. 8 and Fig. 9 
when we remember that there is no fitting parameter
in Eqs. (\ref{perturb-T-offdiag}) and (\ref{SL2}).
As the parameters $w_{\sigma',\sigma}$ shown in Table I
are determined uniquely by Eq. (\ref{def-w}) with 
the interlayer transfer integrals $W$, 
we cannot modify $w_{\sigma',\sigma}$ in order to
improve the agreement.
For the quantitative comparison between Fig. 8 and Fig. 9,
numerical values of $(N, E, T)$ at the local maximums are 
shown in Table III.
Difference between Fig. 8 and Fig. 9 in Table III
is satisfactorily small, though 
the positive (negative) $E$ tends to suppress (enhance) the exact $T$ 
in comparison to Eqs. (\ref{perturb-T-offdiag}) and (\ref{SL2}).
Figure 10 is the same exact calculations as in Fig. 9 except the residue $l=-1$.
The analytical results corresponding to Fig. 10 can be found in Fig. 8
with the transformation (\ref{relation}).
According to the transformation (\ref{relation}),
$T_{-\sigma',-\sigma}$ 
of Fig. 8 is multiplied by $(w_{\sigma',\sigma}/w_{-\sigma',-\sigma})^2$ 
and is shown in the rightmost column of Table III.
It coincides well with $T_{\sigma',\sigma}$ of Fig. 10.
Since $E($ii$)+E($iv$)=\varepsilon$, Eq. (\ref{relation-E}) can be applied
to ii and iv. It follows that
$T($ ii, Fig. 9) $\simeq$ $T($ iv, Fig. 10) and $T($ ii, Fig. 10) $\simeq$ $T($ iv, Fig. 9).
Figures 11 and 12 are the approximate and exact calculations as in Fig. 8
and Fig. 9, respectively, in case $l=0$.
When $l=0$, $N_w=N$ and $T_{\mp,-} = T_{\pm,+}(w_{\mp,-}/w_{\pm,+})^2$ 
in Eq. (\ref{relation}).
It follows that the surfaces $(N,E,T_{\mp,-})$ have the same shape as $(N,E,T_{\pm,+})$ in
Eqs. (\ref{perturb-T-offdiag}) and (\ref{SL2})
, thus we show only $T_{\pm,+}$ in Fig. 11.
Figure 12 also shows that $T_{\pm,+}$ and $T_{\mp,-}$ are similar in shape.
Table IV shows a comparison between Figs. 11 and 12 in the same way
as Table III.
There are eight local maximums labeled by $\bullet$i $\sim$ $\bullet$iv,
($\bullet=\triangle,\Box$) in Fig. 12. In comparison to $\bullet$iii,
the peak $\bullet$iv is very low 
and cannot be visible in Figs. 11 and 12.
The local maximums $\triangle$i $\sim$ $\triangle$iv in Fig. 12
are reproduced properly in Fig. 11.
The heights of $\triangle$ peaks of Fig. 11 are
transformed into those of the $\Box$ peaks by Eq. (\ref{relation}) 
in the rightmost column of Table IV.
Overall, the transformed values
appropriately reproduce the $\Box$ peaks in Fig. 12,
except for the overestimation of the $\Box$iii peak height.

\begin{table}
\begin{tabular}{|c|c|c|c|
} 
$w_{+,+}$ & $w_{-,-}$ & $w_{-,+}$ & $w_{+,-}$ \\
10.78 & $-12.62$ & 
$-9.372 $ & 
$12.62$ 
\end{tabular}
\caption{The parameter $w_{\sigma',\sigma}$ [meV] in case
$(n_\downarrow,n_\uparrow)=(5,10)$ and $(\theta_{\rm d},z_{\rm d})=
(0,0)$. }
\end{table}

\renewcommand{\arraystretch}{1.5}
\begin{table}
\begin{tabular}{|c|c|c|c|c|c|} 
$\varepsilon$ [eV] & 0 & 0.07 & 0.14 & 0.21 & 0.28 \\
\hline
$\frac{1}{2}\lambda_E,\;\frac{3}{2}\lambda_E$ 
& 49.9, 150 & 65.1, 195 & 93.5, * & 166, * & *, * \\
$\frac{1}{6}\lambda_E,\;\frac{7}{6}\lambda_E$ 
& 16.6, 116 & 21.7, 152 & 31.2, * & 55.4, * & *, * \\
$\frac{5}{6}\lambda_E,\;\frac{11}{6}\lambda_E$
& 83.1,$\;$ 183 & 108, *
& 156, * & *, * & *, * \\
\hline
$\lambda_\Gamma$ & * & * & 201 & 139 & 105 \\
\end{tabular}
\caption{The nodes of dashed lines in Fig. 7 .
They are 
$\frac{1}{2}\lambda_E,\;\frac{3}{2}\lambda_E$ $(l=0)$,
$\frac{1}{6}\lambda_E,\;\frac{7}{6}\lambda_E$ $(l=1)$, 
$\frac{5}{6}\lambda_E,\;\frac{11}{6}\lambda_E$ $(l=-1)$
and $\lambda_\Gamma$.
The nodes outside the region $0 < N < 200$ 
are indicated by * except $\lambda_\Gamma =201$. }
\end{table}
\renewcommand{\arraystretch}{1}

\begin{table}
\begin{tabular}{c|c|c|c|c|c|c|c|c|c|c}
&
\multicolumn{3}{c|}{Fig. 8 $(l=1)$} &
\multicolumn{3}{c|}{Fig. 9 $(l=1)$} &
\multicolumn{3}{c|}{Fig. 10 $(l=-1)$} &
\\
\hline
& $N$ & $E$ & $T$ 
& $N$ & $E$ & $T$ 
& $N$ & $E$ & $T$ &
\\ 
i & 49 & $-54$ & 0.078 & 49&$-65$ & 0.088
& 47 & $-62$ & 0.111 & 0.107 \\
ii & 49 & 252 & 0.078 &49& 248& 0.058
& 47 & 262 & 0.091 & 0.107 \\
iii & 49 & $-257$ & 0.104 & 43 & $-297$& 0.121 
& 47 & $-289$ & 0.104 & 0.076 \\
iv & 49 & 48 & 0.104 & 46& 43& 0.099
& 47 & 37 & 0.071 & 0.076 \\
v & 52 & 150 & 0.116 & 49& 147 & 0.122
& 50 & 145 & 0.165 & 0.210 \\
vi & 79 & 150 & 0.598 & 76 & 153 & 0.418
& 77 & 150 & 0.305 & 0.330\\
\end{tabular} 
\caption{The local maximums in Figs. 8, 9 and 10.
The unit of $E$ is meV.
According to the transformation (\ref{relation}),
$T_{-\sigma',-\sigma}$ 
of Fig. 8 is multiplied by $(w_{\sigma',\sigma}/w_{-\sigma',-\sigma})^2$ 
and is compared to $T_{\sigma',\sigma}$ of Fig. 10 
in the rightmost column.}
\end{table}

\begin{table}
\begin{tabular}{c|c|c|c|c|c|c|c|c|c|c}
&
\multicolumn{3}{c|}{Fig. 11 $(\bullet
=\triangle)$ } &
\multicolumn{3}{c|}{Fig. 12 $(\bullet=\triangle)$} &
\multicolumn{3}{c|}{Fig. 12 $(\bullet=\Box)$ } & 
\\
\hline
& $N$ & $E$ & $T$ 
& $N$ & $E$ & $T$ 
& $N$ & $E$ & $T$ &
\\
$\bullet$i & 48 & $-162$ & 0.078 & 
48 & $-173$ & 0.096 &
48 & $-153$ & $0.117$ &0.107\\ 
$\bullet$ii & 48 & $150$ & 0.078 & 
51 & 145 & 0.067 &
51 & 152 & 0.098 & 0.107\\ 
$\bullet$iii & 108 & 150 & 0.675 
& 105 & 151 & 0.520 &
105 & 151 & 0.683 & 1.224\\ 
$\bullet$iv & 27 & 150 & 0.020 
& 27 & 129 & 0.025 &
24 & 165 & 0.027 & 0.036\\ 
\end{tabular}
\caption{
The local maximums of Figs. 11 and 12.
The unit of $E$ is meV.
According to the transformation (\ref{relation}),
$T_{-\sigma',-\sigma}$ 
of Fig. 11 is multiplied by $(w_{\sigma',\sigma}/w_{-\sigma',-\sigma})^2$ 
and is compared to $T_{\sigma',\sigma}$ of Fig. 12 
in the rightmost column.}
\end{table}

The exact dispersion relation near the valley $k = \frac{2\pi}{3a}$
is displayed in Fig. 13 for the discrete energies with the interval 0.002 eV. 
In contrast to the approximate dispersion relation in Fig. 4, the energy
gap appears near the cross point at $E=\varepsilon/2 = 0.15$ eV,
or equivalently, $\widetilde{E} =0$.
In the energy gap, the evanescent waves replace the propagating waves.
Accordingly, the phase factor $e^{ika}$ is replaced
by $e^{(ik \pm \kappa) a}$ in the exact calculation.
The decay factors $e^{-\kappa a}$
are displayed by
the horizontal axes with the vertical axis $E$ in the inset of Fig. 13.
As has been reported by Ref. \cite{slow-decay-side-contact-ref},
the decay length $1/\kappa$ is considerably long.
It suggests the persistence
of the propagating characteristic that enables
Eqs. 
(\ref{perturb-T-offdiag}) and (\ref{SL2})
remain effective in the gap.
Figures 14 and 15 show 
(a) $T_{\sigma,\sigma}$ and (b) $T_{-\sigma,\sigma}$
as a function of $N$ at the center of the gap
$E=\varepsilon/2$.
Black and grey lines correspond to $\sigma= +$ and 
$\sigma= -$, respectively.
As for the residue $l$, $l=1$ in Fig. 14 and $l=0$ in Fig. 15.
We can distinguish the exact results from the approximate ones in the same way as Fig. 7.
Because Eq. (\ref{perturb-T-offdiag}) exceeds unity for the large $N$, 
the dashed lines
of the intervalley $T_{-\sigma,\sigma}$ are limited to the small $N$.
The insets show the magnification of the intervalley $T_{-\sigma,\sigma}$.
The local maximums $\triangle$iv and $\Box$iv in Figs. 11 and 12 can be visible.
Agreement between the solid and dashed lines is
satisfying in the range $N < 50$.
In a wider range of $N$, the solid and dashed lines do not 
necessarily coincide with each other,
while the periods of the solid lines excellently coincide
with that of the dashed lines $2\sqrt{3}t\pi/\varepsilon \simeq 100$.
Though the height ratio between the neighboring peaks in the solid line fluctuates, it approaches $e^{-\kappa 100a} \simeq (0.992)^{100} =0.45$ as $N$ increases.
It comes from the decay characteristic of the gap state.

Figure \ref{Fig-I} displays the product of the squared overlap integrals 
(\ref{def-I}) for the four values of $\varepsilon=$ 0, 0.05, 0.1, 0.15 eV.
Applying Eq. (\ref{def-w}), we obtain $w_{\sigma',\sigma}=\sigma' t_\bot /4$.
As Table I shows $w_{\sigma',\sigma}\simeq 0.01\sigma'$ eV,
$t_\bot$ is set to 0.04 eV.
The black and gray (red online) lines indicate the intervalley $I_{-\sigma,\sigma}$ 
and intravalley $I_{\sigma,\sigma}$, respectively.
They reproduce the qualitative dependence of $T_{\sigma',\sigma}$
on $\varepsilon$ and $E$.
The asymptotic formulas of $I_{\sigma',\sigma}$
are shown
by Table V in case (a) $|\widetilde{E}| \ll |\varepsilon|, |t_\bot|$
and case (b) $|\widetilde{E}| \gg |\varepsilon|, |t_\bot|$
with the parameter $\alpha=\varepsilon/t_\bot$.
Since $|w|=|t_\bot|/4$, the factors $64|w|^2/\Gamma^2$
and $16|w|^2/\widetilde{E}^2$ in Eqs. (\ref{Ysigma}) 
and (\ref{perturb-T-offdiag}) 
are translated to $4/(1+\alpha^2)$ and $|t_\bot/\widetilde{E}|^2$, respectively.
Multiplying them by $1/4$, we get the terms $\delta_{\sigma,\sigma'}/(1+\alpha^2)$ and $\frac{1}{4}\delta_{\sigma,-\sigma'}|t_\bot/\widetilde{E}|^2$ 
in Table V.
Here we can speculate that the factor $1/4$ comes from the average $\lim_{N \rightarrow \infty}\frac{1}{N}\sum_{j=1}^N\sin^2(a'j)
\cos^2(b'j-\phi)= 1/4$ with the constants $a',b'$ and $\phi$.
Figure \ref{Fig-vecD} gives us an illuminating insight into Table V as follows.
Firstly, we discuss case (a) 
where
Eq. (\ref{overlap-def-AB}) is approximated by
$q_{\rm r}=\tau'\alpha/\sqrt{\alpha^2+2i\tau|\alpha|}$ and $q_\uparrow=
\pm i $.
The left-to-right amplitude ratio
is 1 to $q_{\rm r}$ in ladder $\downarrow$, while
1 to $-q_{\rm r} $ in ladder $\uparrow$ 
because $E'/E=-1$.
As the ladders $\downarrow$ and $\uparrow$
are nearly opposite in the symmetry in this way,
$I_{-\sigma,\sigma} > I_{\sigma,\sigma}$
in case (a). 
When $\alpha$ approaches zero,
however, $q_{\rm r}$ also approaches zero.
The limit is the average of a symmetric state and an antisymmetric one.
Accordingly $I_{-\sigma,\sigma} = I_{\sigma,\sigma}$ when $\varepsilon=
\widetilde{E}=0$.
Next Let us discuss case (b) where $q_{\rm r} \simeq \tau' \frac{E}{|E|} = \pm 1, q_\uparrow\simeq \alpha-\tau\frac{E}{|E|}\sqrt{1+\alpha^2}$.
The two ladders have almost the same left-to-right amplitude ratio
since $E'/E \simeq 1$. 
Therefore $I_{-\sigma,\sigma}$ is much less than $ I_{\sigma,\sigma}$.
The small nonzero $I_{-\sigma,\sigma}$ being inversely
proportional to $\widetilde{E}^2$ originates from the slight difference of $|q_{\rm r}|$ from 1 as is explained
by Appendix D.
In order to understand the term $\delta_{\sigma',\sigma}/(1+\alpha^2)$ 
in Table V, we should refer to the parameter $|q_\uparrow|$ that represents the interladder localization.
As $\alpha$ approaches zero, $|q_\uparrow|$ tends to 1.
It leads to the extended state between the two ladders 
and the large $I_{\sigma,\sigma}$.
When $|\alpha| \gg 1$, on the other hand, $|q_\uparrow|$ is close to $(2|\alpha|)^{\pm 1}$.
It causes the localized state at the single ladder
and the small $I_{\sigma,\sigma}$ .
\renewcommand{\arraystretch}{2}
\begin{table}
\begin{tabular}{c|c|c}
& (a) $|\widetilde{E}| \ll |\varepsilon|,|t_\bot|$ & (b) $|\widetilde{E}| \gg |\varepsilon|,|t_\bot|$ \\
arbitrary $\alpha$ & $\frac{1}{2}-\frac{\sigma\sigma'|\alpha|}{|\alpha|+\sqrt{\alpha^2+1}}$ 
& $\frac{\delta_{\sigma,\sigma'}}
{1+\alpha^2}+\delta_{\sigma,-\sigma'}
\frac{2\alpha^2+1}{8\alpha^2+8} t_\bot^2\widetilde{E}^{\;-2}
$ \\
$ \alpha=0$ & $\frac{1}{2}$ & $\delta_{\sigma,\sigma'}+\delta_{\sigma,-\sigma'}\frac{1}{8}
t_\bot^2\widetilde{E}^{\;-2}$
\\
$\alpha=\infty$ & $
\delta_{\sigma,-\sigma'}$ & $\delta_{\sigma,-\sigma'}\frac{1}{4} t_\bot^2\widetilde{E}^{\;-2}$
\end{tabular}
\caption{Asymptotic formulas of Eq. (\protect\ref{def-I}) with the parameter
$\alpha=\varepsilon/t_\bot$ 
}
\end{table}
\renewcommand{\arraystretch}{1}
As examples of the influence of $\theta_{\rm d}$ and $z_{\rm d}$,
Fig. 17 shows
the conductance $G=\sum_{\sigma,\sigma'}T_{\sigma',\sigma}$ 
in the unit of $2e^2/h$ as a function of the energy $E$ 
in case $(\theta_{\rm d},z_{\rm d})=(0,\pm a/20),(-\pi/50,0),(\pi/30,0)$, $N=$ 101, 102, 103.
The black and gray lines correspond to $\theta_{\rm d}=0$ and $z_{\rm d} =0$,
respectively.
The dashed (solid) lines show 
the approximate (exact) results.
Though $w_{\sigma'\sigma}$ becomes different from that of Table I, the condition $|w_{\sigma',\sigma}| \ll |\varepsilon|$ remains valid.
Thus the resonant peaks of the intervalley terms $T_{+,-}, T_{-,+}$ remain dominant.
We can see that the center of the peak, $E=\varepsilon/2$,
is insensitive to $N$, $\theta_{\rm d}$ and $z_{\rm d}$.
When the $\theta_{\rm d}=-\pi/30$, the mirror planes of the two tubes
coincide with each other, and thus $w_{+,-}=w_{-,+}=0$.
It explains why the conductance of $\theta_{\rm d}= -\pi/50$
is remarkably smaller than the other $\theta_{\rm d}$.
Except the case $\theta_{\rm d}= -\pi/50$ where
the agreement between the solid and dashed lines is fairy good, the dashed lines show overestimated values of $G$.
Nevertheless the resonant intervalley peak survives in all the solid lines.
Equation (\ref{tuika}) enables us to approximate
the difference $\Delta G \equiv G\left(\frac{a}{20}\right)-G\left(\frac{-a}{20}\right)
$ for the black dashed lines of Fig. 17
as $\frac{a}{10}\sum_\sigma C_\sigma\sin\left(-\frac{2}{3}\pi\sigma N \right)
\frac{\partial \phi_{-\sigma,\sigma}}{\partial z_{\rm d}} $ 
where $\varepsilon N/(\sqrt{3}t) \simeq 2\pi$.
Since $C_+ \simeq C_- >0$ and 
$\frac{\partial \phi_{+,-}}{\partial z_{\rm d}} -\frac{\partial \phi_{-,+}}{\partial z_{\rm d}} >0$,
$\Delta G$ has the same sign as $\sin(2\pi N/3)$, i. e., 
$\Delta G(N=101) < 0$, $\Delta G(102) \simeq 0$,
and $\Delta G(103) > 0$.
This result is consistent with the solid black lines 
indicating that the effect of $z_{\rm d}$ is
appropriately included by Eq. (\ref{perturb-T-offdiag}).

Some peaks in Fig. 14 (b) and Fig. 15 (b) 
approximately satisfy
the conditions of the VCR and the VCF discussed in Sec. III.
The peak indicated by $\triangleright$ ($\triangleleft$)
in Fig. 14 (b) 
works not only as the VCR for the positive (neative) pure $J{\rm (L)}_v$
but also as the VCF producing the $\psi{(\rm R)}_{K',\rightarrow}$ 
( $\psi{(\rm R)}_{K,\rightarrow}$ ).
On the other hand, the peak indicated by $\circ$ 
in Fig. 15 (b) approximately satisfies both (i) $T_{K',K}=1$ 
and (ii) $T_{K,K'}=1$ ,
indicating the VCR irrespective of the sign of the pure $J{(\rm L)}_v$,
while it does not work as the VCF.
Since the GrS and the po-Gr of Refs. 
\cite{graphene-precession,gra-gra-junction}
must obey the condition $T_{\sigma',\sigma}=T_{-\sigma',-\sigma}$ that conflicts with the condition of the VCF,
they cannot be the VCF.
Contrary to it, Eq. (\ref{relation}) explicitly shows that
$T_{\sigma',\sigma} \neq T_{-\sigma',-\sigma}$
for the ps-ANT.
Note that the time reversal symmetry
is {\it no} guarantee of the condition $T_{\sigma',\sigma} = T_{-\sigma',-\sigma}$.

The nonzero terms in Eq. (\ref{def-w}) 
are limited to the sites $\theta_{j,m} \simeq 0$ in
the ps-ANTs, whereas they spread over
the all $m$ in the t-ANTs.
It follows that
$w_{+,+} $ of the t-ANTs is comparable to $t_1(=0.36$ eV)
being much larger than $w_{+,+}$ of the ps-ANTs.
As it is difficult to make
$|\varepsilon|$ much larger than $t_1$,
the ps-ANT is more suitable than the t-ANT
for the condition $|\varepsilon| \gg w_{+,+}$.
The perturbative calculation is effective
on condition that $|E|, |\varepsilon|, |w_{\sigma',\sigma}| 
\ll t$.
The condition $ |w_{\sigma',\sigma}| \ll t$ always holds.
The other conditions $|E|, |\varepsilon| \ll t$
are also satisfied except when
the applied gate voltage and the doping strength 
are extremely high.
Author has confirmed the approximate formulas
(\ref{perturb-T-offdiag}) and (\ref{SL2})
reproduce the exact results
irrespective of the choice of $n_\xi$ (not shown in Figures).
The effect of $n_\xi$ on 
Eqs. (\ref{perturb-T-offdiag}) and (\ref{SL2}) can be derived from Eq. (\ref{def-w})
that is the explicit relation between $w_{\sigma',\sigma}$ 
and the interlayer Hamiltonian $W$.
Though various atoms and molecules
can be encapsulated, some of them might
be unsuitable for the VCR and VCF.
When europium and potassium 
atoms are encapsulated, for example,
the current via the metal nanowire \cite{Eu-wire} and
the nearly free electron states \cite{NearlyFree} 
might hide the VCR and the VCF.
For more detailed discussions, 
the first principle calculation (FP) is desirable.
To apply Eq. (\ref{def-w2}) to
the FP, we only have to
replace the wave function $\vec{b}$ and the 
interlayer Hamiltonian elements $W$
with those of the FP. \cite{tamura-2012}

\section{summary and conclusion}
The interlayer transmission rates 
of the ps-ANTs
 have been calculated by the $\pi$ orbital
TB with the intertube site energy 
difference $\varepsilon$.
The valley channels 
can be interpreted as the symmetric $(+)$ 
and antisymmetric $(-)$ channels concerning
the plane including the tube axis.
Considering the interlayer transfer integral to be the perturbation,
we have derived the approximate analytical formulas
(\ref{perturb-T-diag}) and (\ref{perturb-T-offdiag}) 
that is determined by the electron energy $E$, the integer overlap length $N$
and the interlayer Hamiltonian element $w_{\sigma',\sigma}$ 
between $\sigma'$ and $\sigma$ channels.
The geometrical overlap length equals $(N-2)\frac{a}{2}-z_{\rm d}$
with the lattice constant $a=$ 0.246 nm and the small translation $|z_{\rm d}| <\frac{a}{4}$.
Equation (\ref{def-w}) enables us to
transform the TB interlayer transfer integrals into $w_{\sigma',\sigma}$
without a fitting parameter.
The effect of $z_{\rm d}$ is included by $w_{\sigma',\sigma}$.

In comparison to Ref.\cite{tamura-2019} ,
the nonzero $\varepsilon$
requires
only a slight difference in the numerical code of
the exact calculation.
The perturbative
calculations, however,
have to be performed
in a qualitatively different
way since
the nonzero $\varepsilon$ lifts
the degeneracy of the unperturbed system.
As $|\varepsilon|$ deceases, the
system approaches the degenerate case
and the perturbation theory for the nondegenerate case
becomes ineffective.
It is the reason why Eq. (\ref{perturb-T-diag}) with zero $\varepsilon$ does not reproduce the formula 
in Ref.\cite{tamura-2019} .
Therefore it is not a trivial result
that Eq. (\ref{perturb-T-offdiag}) with zero $\varepsilon$ 
is identical with
the formula in Ref. \cite{tamura-2019}.
Equation (\ref{SL2}) is an interpolation 
between the degenerate case $(\varepsilon=0$, Ref. \cite{tamura-2019} )
and the nondegenerate case Eq. (\ref{perturb-T-diag}).
The effective range of
Eqs. (\ref{perturb-T-offdiag}) 
and (\ref{SL2})
with respect to $E$ and $N$ is satisfyingly large.
Tables III and IV are the guides for the quantitative comparison
between the exact results
and the approximate analytical formulas.
They afford an archetypal example of
the effectiveness of the analytical formulas.

The product of the squared overlap integrals (\ref{def-I}) 
has a close relation to the wave function and explains the averaged
transmission rate $
\frac{1}{N} \sum_{j=1}^NT(N)$ with sufficiently large $N$.
Figure \ref{Fig-vecD} and Eq. (\ref{overlap-vecD}) clearly indicate 
that the VCR
$T_{-\sigma,\sigma} \simeq 1 \gg T_{\sigma,\sigma}$ occurs when 
$|q_\uparrow| \simeq |q_{\rm r}|\simeq 1$ and $E'/E \simeq -1$, i.e., on condition that $\widetilde{E}=0$ and $|\varepsilon| \gg |t_\bot|
(\simeq 4|w|)$.
In order to interpret the dependence on $N$, however,
Eq. (\ref{def-I}) is ineffective while
Eqs. (\ref{perturb-T-offdiag}) 
and (\ref{SL2}) are necessary.
The peak at $E=\varepsilon/2$
in the $E$-$T_{-\sigma,\sigma}$ curve 
survives
the change of the interlayer configuration $(\theta_{\rm d},z_{\rm d})$ 
if the two mirror planes are not very close to 
each other.
At the peak energy $E=\varepsilon/2$, the oscillation of
the exact $N$-$T_{-\sigma,\sigma}$ curve
grows into about unit first,
then decays exponentially.
The factor of the corresponding analytical formula 
$N^2\cos^2\left(\frac{\varepsilon N}{2\sqrt{3}t}
\pm \frac{\pi}{3}N \right)$
might be the origin of the former growing oscillation.

The ps-ANT works both as the VCR and as the VCF,
for example, peaks indicated by $\triangleleft,
\triangleright, \circ$ in Figs. 14 (b) and 15 (b).
The resonant peak in the $E$-$G$ curve
corresponding to the VCR and the VCF
has two distinguishing characteristics
not found in other double junction systems.
Firstly the peak energy $E=\varepsilon/2$ 
is not affected by the overlap length.
In the experiment, 
the mechanical motion of the piezo electrode attached to each ANT
enables us to modulate the overlap length 
with other parameters unchanged.
Secondly, it is evident from Eq. (\ref{perturb-T-offdiag}) that the peak at $E=\varepsilon/2$ is dominant over the other peaks.
In contrast, multiple peaks have similar heights in the usual $E$-$G$ curve.
These characteristics are fingerprints of the VCR and the VCF.

\appendix

\section{Derivation of Eqs. (\ref{SL}) and (\ref{SR})}
The explicit formulas of Eq. (\ref{def-X})
are 
\begin{equation}
X^{[0]}_{\rm L}
=V_0
\left(
\begin{array}{ccc}
\Omega
,& 0,
& -\Omega^*
\\
{\bf 1}_2,& 0
, & -{\bf 1}_2\\
0,& {\bf 1}_2
& 0\\
\end{array}
\right),
\label{XL0} 
\end{equation}
\begin{eqnarray}
X^{[1]}_{\rm L}
&=& 
V_0
\left(
\begin{array}{ccc}
0,& F_-\Omega
& 0\\
0,& F_-
, & 0\\
\;^tF_+
,& 0
& 0\\
\end{array}
\right),
\label{XL1} 
\end{eqnarray}
\begin{equation}
X^{[0]}_{\rm R}
=
V_0
\left(
\begin{array}{ccc}
0,& \Omega
& -\Omega^*
\\
0,& {\bf 1}_2,
& -{\bf 1}_2\\
{\bf 1}_2,
& 0, & 0\\
\end{array}
\right),
\label{XR0}
\end{equation}
\begin{eqnarray}
X^{[1]}_{\rm R}
&=& 
V_0
\left(
\begin{array}{ccc}
\;^tF_+^*\Omega,&0,
& 0\\
\;^tF_+^*,& 0, & 0\\
0,& F_-^*,
& 0\\
\end{array}
\right)
\label{XR1} 
\end{eqnarray}
with the notation
\begin{equation}
V_0
=\left(
\begin{array}{ccc}
v_0
,& 0,
& 0
\\
0,& v_0
, & 0\\
0,& 0,
& v_0\\
\end{array}
\right)
\label{App-V0} 
\end{equation}
Inverse matrixes of Eqs.(\ref{XL0}) and (\ref{XR0} ) are represented by
\begin{equation}
\left(X^{[0]}_{\rm L}\right)^{-1}
=
\frac{-i}{\sqrt{3}}
\left(
\begin{array}{ccc}
\sigma_z
,& 
-\Omega^*\sigma_z,& 
0\\
0,& 0
, & i\sqrt{3} \\
\sigma_z

,& 
-\Omega\sigma_z
,& 
0\\
\end{array}
\right)V_0^{-1}
\label{inverse-XL} 
\end{equation}

\begin{equation}
\left(X^{[0]}_{\rm R}\right)^{-1}
=
\frac{-i}{\sqrt{3}}
\left(
\begin{array}{ccc}
0,& 0
, & i\sqrt{3} \\
\sigma_z
,& 
-\Omega^*\sigma_z,& 
0\\
\sigma_z

,& 
-\Omega\sigma_z
,& 
0\\
\end{array}
\right)V_0^{-1}
\label{inverse-XR} 
\end{equation}
with the Pauli matrix $\sigma_z$.
The following is the perturbation series of
$S_\mu=-X_\mu^{-1}X_\mu^*$ 
up to the first order.
\begin{eqnarray}
S_\mu
&=& 
-\left(X_\mu^{[0]}\right)^{-1}
X_\mu^{[0]*}
\nonumber \\
&& +\left(X_\mu^{[0]}\right)^{-1}
\left[ X_\mu^{[1]}\left(X_\mu^{[0]}\right)^{-1}
X_\mu^{[0]*}
-X_\mu^{[1]*}\right]
\end{eqnarray}
From these equations, we obtain Eqs. (\ref{SL}) and (\ref{SR}).

\section{
derivation of Eq. (\ref{SL2}) in the absence
of the intervalley scattering }

When we neglect the intervalley parameters $2w_{\pm,\mp}$,
the eigen value equation $H(k)\vec{b}=E\vec{b}$ 
for symmetry $\sigma$ becomes 
\begin{equation}
\left(
\begin{array}{cc}
E-E_{\sigma,\downarrow}, & -2w_{\sigma,\sigma}
\\-2w_{\sigma,\sigma}^*, &E-E_{\sigma,\downarrow}-\varepsilon
\end{array}
\right)
\left(
\begin{array}{c}
d_{\sigma,\downarrow}' \\
d_{\sigma,\uparrow }'
\end{array}
\right)
=
\left(
\begin{array}{c}
0 \\
0
\end{array}
\right)
\label{ladder-TB} 
\end{equation}
where $w_{\sigma,\sigma}=w_{\sigma,\sigma}^{(\sigma)}$ and $E_{\sigma,\downarrow}$ is determined by Eq. (\ref{disper-0}).
We derive the dispersion relation 
\begin{equation}
E=\sigma t \left[ 2\cos\left(\frac{k_m}{2}
a\right) - 1 \right] +\frac{\varepsilon+(-1)^m\Gamma_\sigma}{2}
\label{ladder-E}
\end{equation}
from Eq. (\ref{ladder-TB}) 
where $m=1,2$.
Whereas the first order shift of the energy is zero in Sec. \ref{sec-perturb}, 
Eq. (\ref{ladder-E})
depends on the interlayer element $2w_{\sigma,\sigma}$.
Here neglect of the intervalley $w_{\mp,\pm}$ is the tradeoff
for the nonzero energy shift caused by the intravalley $w_{\sigma,\sigma}$.
The wave function of region D is
represented by
\begin{equation}
\left(
\begin{array}{c}
c_{j,\downarrow}^{\;\prime(\rm D)}
\\
c_{j,\uparrow}^{\;\prime(\rm D)}
\end{array}
\right)
=
\sum_{s=\pm}
\left(
\begin{array}{cc}
c_\sigma^{(s)}
,\;& -s_\sigma^{(s)} \\
s_\sigma^{(-s)},\;& c_\sigma^{(-s)} 
\end{array}
\right)
\Xi_\sigma^{s(j+1)}\vec{\gamma}_s^{\;(\rm D)}
\label{ladder-cjD} 
\end{equation}
where
\begin{equation}
(c_\sigma^{(\pm)}, \;s_\sigma^{(\pm)})
=
e^{\pm i\frac{\phi_{\sigma,\sigma}}{2}} \left(\sqrt{\frac{\varepsilon+\Gamma_\sigma}{2\Gamma_\sigma}},
\;
-\sqrt{\frac{-\varepsilon+\Gamma_\sigma}{2\Gamma_\sigma}} \right)
\label{ladder-cs} 
\end{equation}
and $(\Xi_\sigma)_{m,m'}=- \delta_{m,m'}\exp(ik_ma/2)$.
In the following, we use abbreviations $c_\sigma=c_\sigma^{(+)}$ 
and $s_\sigma=s_\sigma^{(+)}$.
When $k_m \simeq -2\pi\sigma/(3a)$, $\Xi_\sigma$ is approximated by
\begin{equation}
\Xi_\sigma
=-\exp(i\eta_{\sigma,\sigma}^{(+)})
\left(
\begin{array}{cc}
e^{i\frac{\Gamma_\sigma}{2\sqrt{3}t}}
,& 
0
\\
0,
& 
e^{-i\frac{\Gamma_\sigma}{2\sqrt{3}t}}
\end{array}
\right)\;\;.
\label{ladder-Xi} 
\end{equation}
By the replacement
$U_\mu^{[n]} \rightarrow (c_\sigma,-s_\sigma),$ $ U_{-\mu}^{[n]} \rightarrow 
(s^*_\sigma,c^*_\sigma)$, $v_0 \rightarrow 1$, 
$ \Omega^*\delta_{n,0} \rightarrow 
\exp(-2i\sigma\pi/3)$ and $\Xi_0 \rightarrow \Xi_\sigma
\rightarrow \exp(2i\sigma \pi/3)$ in Eq. (\ref{def-X}), 
we can obtain 
\begin{equation}
X_{\sigma,\rm L}
=\left(
\begin{array}{ccc}
c_\sigma e^{i\sigma\frac{2}{3}\pi}
,& 
-s_\sigma e^{i\sigma\frac{2}{3}\pi}
,& 
-e^{-i\sigma\frac{2}{3}\pi}
\\
c_\sigma,& 
-s_\sigma
,& 
-1
\\
s_\sigma^*
,& 
c_\sigma^*
,& 
0
\end{array}
\right)
\label{ladder-XL}
\end{equation}
where $\widetilde{E}/t$ and $\Gamma_\sigma/t$ are approximated by zero.
We can easily calculate $S_{\sigma,\rm L}=-X_{\sigma,\rm L}^{-1}X_{\sigma,\rm L}^*$ as
\begin{equation}
S_{\sigma,\rm L}
=\left(
\begin{array}{ccc}
-s_\sigma^2
,& 
-c_\sigma s_\sigma
,& 
c_\sigma^*
\\
-c_\sigma s_\sigma
,& 
-c_\sigma^2
,& 
-s_\sigma^*
\\
c_\sigma^*
,& 
-s_\sigma^*
,& 
0
\end{array}
\right)
\label{ladder-SL}
\end{equation}
Replacing $c_\sigma$ and $s_\sigma$ by
$s_\sigma$ and $-c_\sigma$, respectively, 
we also obtain $X_{\sigma,\rm R}$ and $S_{\sigma,\rm R}$.
Using Eqs. (\ref{ladder-Xi}) , (\ref{ladder-SL}) and $S_{\sigma,{\rm R}}$ in Eq. (\ref{tRL}),
we can derive Eq. (\ref{SL2}).

\section{The Upper limit of $I_{\sigma',\sigma}$ and $I_{\sigma,\sigma}+
I_{-\sigma,\sigma}$ }
With the notations
\begin{equation}
(a_1, a_2, a_3)=\left( \frac{E}{\bar{p}_\tau},\;\frac{E^2-\bar{p}_\tau^2}{t_\bot \bar{p}_\tau},\;\frac{E^2-\bar{p}_\tau^2}{t_\bot E'} \right)
\end{equation}
and
\begin{equation}
\bar{p}_\tau=\sqrt{\widetilde{E}^2+\frac{\varepsilon^2}{4}+\tau\sqrt{(\varepsilon^2+t_\bot^2)\widetilde{E}^2-\frac{\varepsilon^2t_\bot^2}{4}}}\;,
\label{overlap-def-q}
\end{equation}
Eq. (\ref{overlap-vecD}) is represented by
\begin{equation}
\;^t\vec{d}_{\rm D}=(1,\;\tau'a_1,\;\tau'a_2,\;a_3)\;.
\end{equation}
Note that $a_1,a_2$ and $a_3$ do not depend on
$\tau'$.
Equation (\ref{def-I}) is represented by
\begin{eqnarray}
I_{\sigma',\sigma} &=& \frac{1}{2}\sum_{\tau'=\pm}\sum_{\tau=\pm}
\frac{\left| 1+\sigma\tau'a_1 \right|^2\left| a_2+\sigma'\tau'a_3 \right|^2}
{\left|1+|a_1|^2+|a_3|^2+|a_4|^2 \right|^2}
\nonumber \\
&=& Q^{(1)}_++Q^{(1)}_-+\sigma\sigma'(Q^{(2)}_++Q^{(2)}_-)
\label{overlap-QR}
\end{eqnarray}
where
\begin{equation}
Q^{(1)}_\tau=\frac{a_4a_5}{(a_4+a_5)^2} \leq \frac{1}{4}
\label{overlap-defQ}
\end{equation}
\begin{equation}
Q^{(2)}_\tau=\frac{(a_1+a_1^*)(a_2a_3^*+a_3a_2^*)}
{(a_4+a_5)^2}
\label{overlap-defR}
\end{equation}
with the notations $a_4=1+|a_1|^2,\;a_5=|a_2|^2+|a_3|^2$.
Formulas (\ref{overlap-QR}) and (\ref{overlap-defQ}) enable us 
to derive
\begin{equation}
I_{\sigma,\sigma}+I_{-\sigma,\sigma}= 2(Q^{(1)}_++Q^{(1)}_-) \leq 1\;.
\label{overlap-QQRR}
\end{equation}
Since $I_{\sigma,\sigma} \geq 0$
 and $I_{-\sigma,\sigma} \geq 0$, Eq. (\ref{overlap-QQRR}) indicates that $I_{\sigma,\sigma} \leq 1$
 and $I_{-\sigma,\sigma} \leq 1$.

\section{The asymptotic formula of $I_{-\sigma,\sigma}$ when
$|\widetilde{E}| \rightarrow \infty$ }
Since
\begin{equation}
\lim_{|\widetilde{E}|\rightarrow \infty}
\frac{q_\uparrow^2}{2\left|\vec{d}_{\rm D}\right|^4}=\frac{1}{32(\alpha^2+1)}
\label{overlap2-1}
\end{equation} 
with the parameter $\alpha=\varepsilon/t_\bot$, 
\begin{equation}
\lim_{|\widetilde{E}|\rightarrow \infty}
I_{-\sigma,\sigma}= \lim_{|\widetilde{E}|\rightarrow \infty}
\sum_{\tau=\pm}\sum_{\tau'=\pm} \frac{Q^{(3)}}
{32(\alpha^2+1)}
\label{overlap2-2}
\end{equation}
where
\begin{eqnarray}
Q^{(3)} &=& 
(1+\sigma q_{\rm r})^2\left(1-\sigma q_{\rm r}\frac{E-\varepsilon}{E}\right)^2 \nonumber \\
&=& 
\left( 1+\frac{\frac{\varepsilon^2}{4}-\widetilde{E}^2}{\bar{p}_\tau^2}
+
\frac{\sigma\tau'\varepsilon}{\bar{p}_\tau}\right)^2
\label{overlap2-3}
\end{eqnarray}
with $\bar{p}_\tau$ defined by Eq. (\ref{overlap-def-q}).
Applying Eq. (\ref{overlap2-3}) to Eq. (\ref{overlap2-2}),
we obtain
\begin{equation}
\lim_{|\widetilde{E}|\rightarrow \infty}
I_{-\sigma,\sigma}=
\frac{2\alpha^2+1}{8(\alpha^2+1)}\left(\frac{t_\bot}{\widetilde{E}}\right)^2
\end{equation} 
since
\begin{equation}
\lim_{|\widetilde{E}|\rightarrow \infty}
\frac{1}{\bar{p}_\tau}=
\frac{1}{|\widetilde{E}|}
\end{equation} 
and
\begin{equation}
\lim_{|\widetilde{E}|\rightarrow \infty}
\frac{\widetilde{E}^2}{\bar{p}_\tau^2}=
1 -\tau\frac{\sqrt{\varepsilon^2+t_\bot^2}}{|\widetilde{E}|}\;\;.
\end{equation}

\begin{figure}
\epsfxsize=\columnwidth
\centerline{\hbox{ \epsffile{newFig1-b.eps} }}
\caption{
Schematic diagrams of (a) the parallel side contacted nanotube (ps-NT)
and (b) the telescoped NT (t-NT).
The two NTs overlap partially.
The right panels are the cross-sectional views of the overlap regions
where the dotted lines indicate the interlayer regions.
The radius difference of the t-NT is close to the interlayer distance
of the graphite.
The gray rectangular are the source and drain electrodes.
The black bar within the NT represents
densely encapsulated iodine atoms. 
}
\end{figure}

\begin{figure}
\epsfxsize=\columnwidth
\centerline{\hbox{ \epsffile{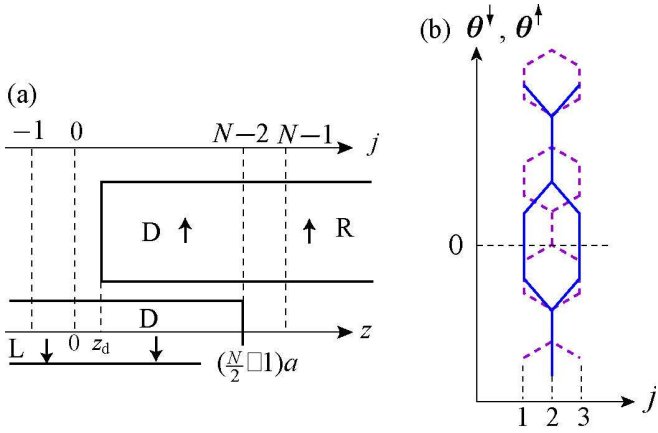} }}
\caption{(a) The spatial configuration of the ps-ANTs (b) The interlayer
configuration when $(\theta_{\rm d},z_{\rm d})=(0,0)$ and
$n_\uparrow=2n_\downarrow$. }
\end{figure}

\begin{figure}
\epsfxsize=\columnwidth
\centerline{\hbox{ \epsffile{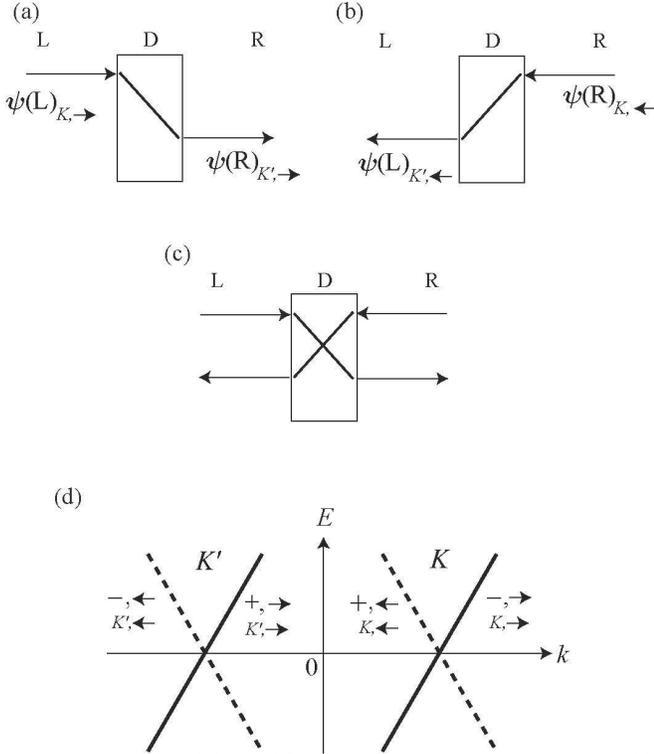} }}
\caption{
(a), (b) , (c) Transmission on condition (i) $T_{K',K}=1$
where $\psi{(\mu)}_{\sharp,\rightarrow}$
( $\psi{(\mu)}_{\sharp,\leftarrow}$) 
denotes the Bloch wave function of region $\mu (=$ L,R)
and of valley $\sharp (= K,K')$ with positive
(negative) $\frac{dE}{dk}$. 
(a) $\psi{\rm (L)}_{K,\rightarrow}$ 
is incident.
(b) $\psi{\rm (R)}_{K,\leftarrow}$ 
is incident.
(c) $\psi{\rm (L)}_{K,\rightarrow}$ and
$\psi{\rm (R)}_{K,\leftarrow}$ are incident at the same time.
(d)
The dispersion relation near the valleys $K$ and $K'$. 
The right $(\rightarrow)$ and left $(\leftarrow)$ going waves 
correspond to the solid and dotted dispersion lines, respectively.
}
\end{figure}

\begin{figure}
\epsfxsize=\columnwidth
\centerline{\hbox{ \epsffile{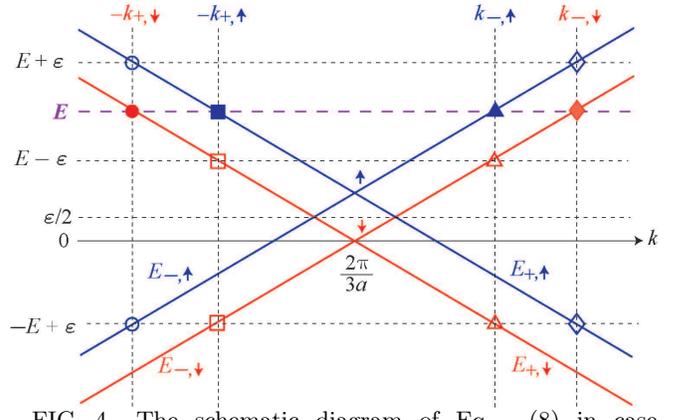} }}
\caption{The schematic diagram of Eq. (\ref{k-E}) in case $k \simeq 2\pi/(3a)$.}
\end{figure}

\begin{figure}
\epsfxsize=.9\columnwidth
\centerline{\hbox{ \epsffile{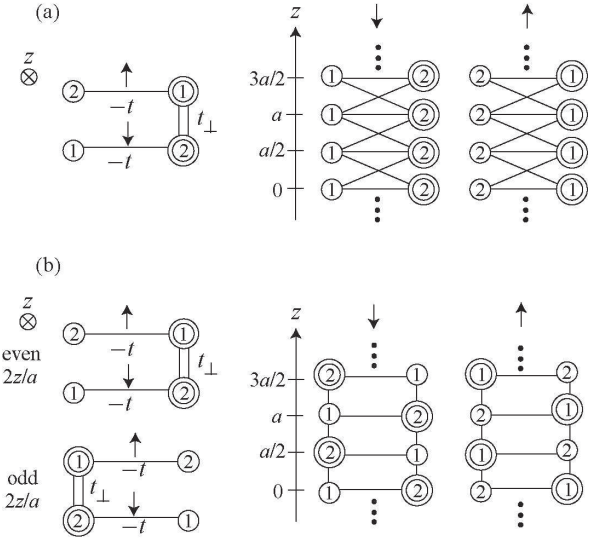} }}
\caption{Two diagrams of the Hamiltonian. 
(a) twisted double ladder with the periodic
interladder transfer integral $t_\bot$. (b) double ladder with staggered $t_\bot$.
}
\label{Fig-ladder}
\end{figure}

\begin{figure}
\epsfxsize=0.4\columnwidth
\centerline{\hbox{ \epsffile{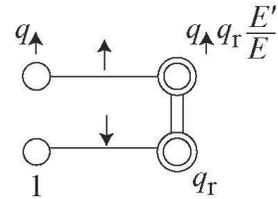} }}
\caption{Relation between Fig. \protect\ref{Fig-ladder} and Eq. (\protect\ref{overlap-vecD}). }
\label{Fig-vecD} 
\end{figure}

\begin{figure}
\epsfxsize=\columnwidth
\centerline{\hbox{ \epsffile{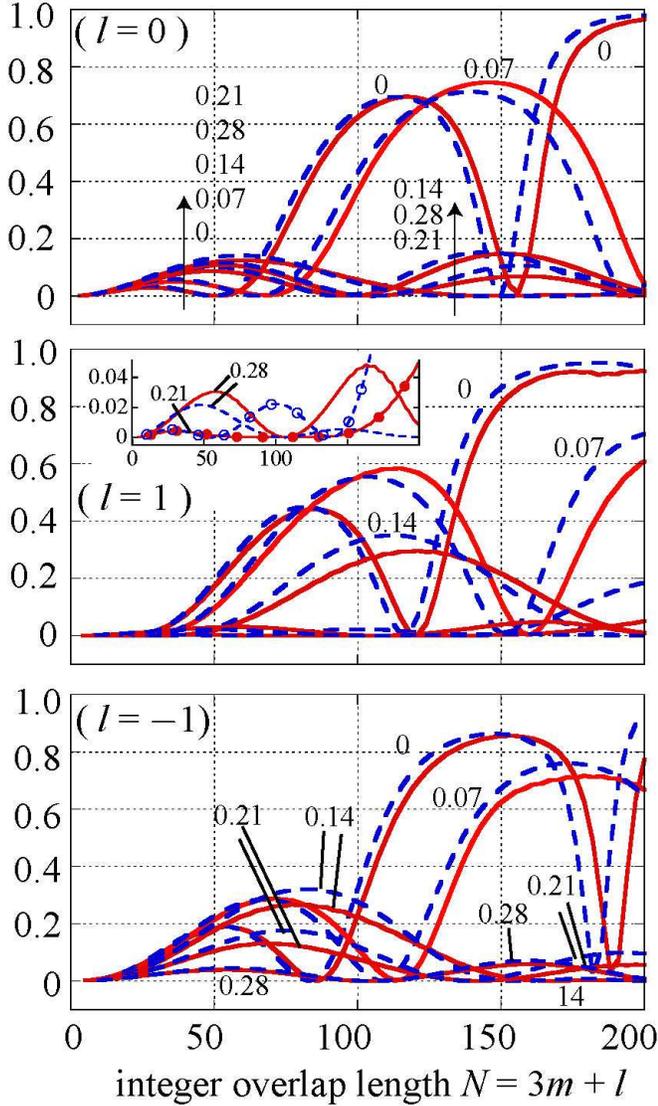} }}
\caption{
$T_{-,-}$ as a function of the integer overlap length $N$
for the energy $E=0.15$ eV
and for the site energy difference
$\varepsilon=$ 0, 0.07,,0.14, 0.21 and 0.28 eV.
The solid and dashed lines display
the exact results and Eq. (\protect\ref{SL2}), respectively.
In each panel, 
$N=3m+l$ of which the integer $m$ is varied while
the residue $l$ is fixed to 0, 1 and $-1$.
The nodes of the dashed lines are listed in Table II.
}
\end{figure}

\begin{figure}
\epsfxsize=\columnwidth
\centerline{\hbox{ \epsffile{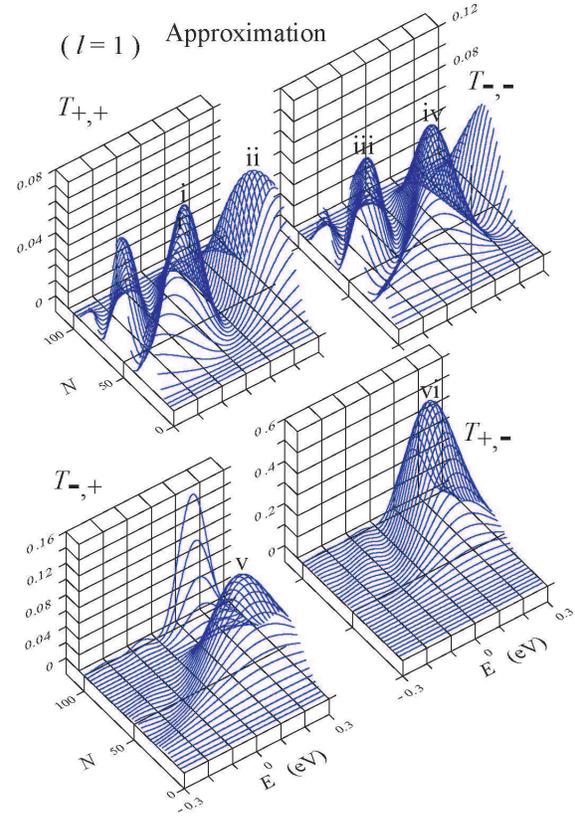} }}
\caption{
Three dimensional view of
Eqs. (\ref{perturb-T-offdiag}) and (\ref{SL2}) in case $l=1$,
$\varepsilon=$ 0.3 eV, $\theta_{\rm d}=0$ and
$z_{\rm d}=0$.
}
\end{figure}

\newpage

\begin{figure}
\epsfxsize=\columnwidth
\centerline{\hbox{ \epsffile{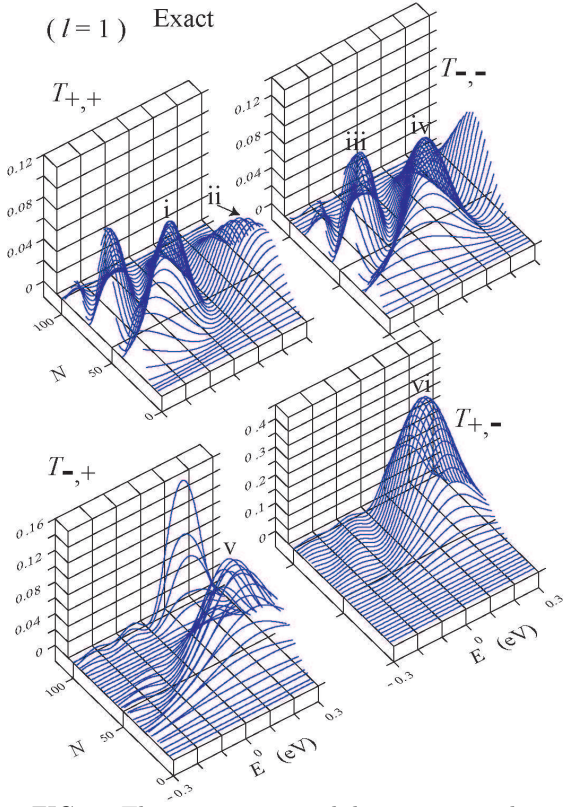} }}
\caption{ The exact numerical data corresponding to Fig. 8.
The local maximums are compared with those of Fig. 8 
in Table III. 
}
\end{figure}

\newpage

\begin{figure}
\epsfxsize=\columnwidth
\centerline{\hbox{ \epsffile{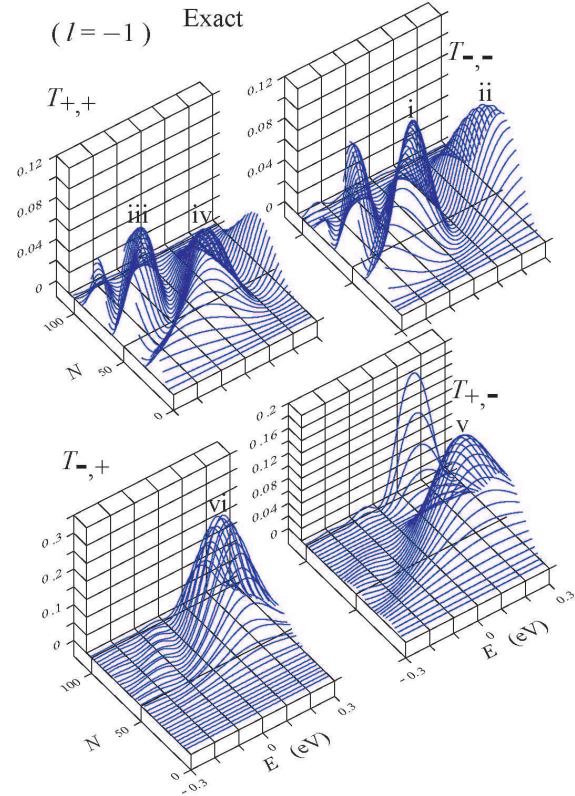} }}
\caption{
The same numerical data as in Fig. 9 except the residue $l=-1$.
The local maximums are compared with those of Fig. 8 
in Table III. The data $T$ of Fig. 8 are transformed by Eq. (\protect\ref{relation}) and shown in the rightmost column of Table III.
}
\end{figure}


\begin{figure}
\epsfxsize=\columnwidth
\centerline{\hbox{ \epsffile{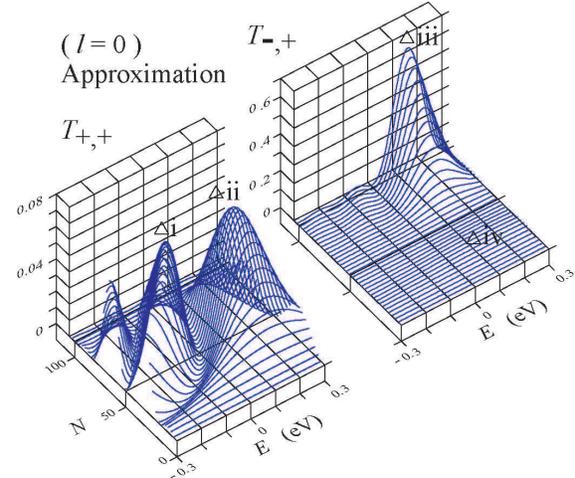} }}
\caption{Three dimensional view of
Eqs. (\ref{perturb-T-offdiag}) and (\ref{SL2}) in case $l=0$,
$\varepsilon=$ 0.3 eV, $\theta_{\rm d}=0$ and
$z_{\rm d}=0$. }
\end{figure}

\begin{figure}
\epsfxsize=\columnwidth
\centerline{\hbox{ \epsffile{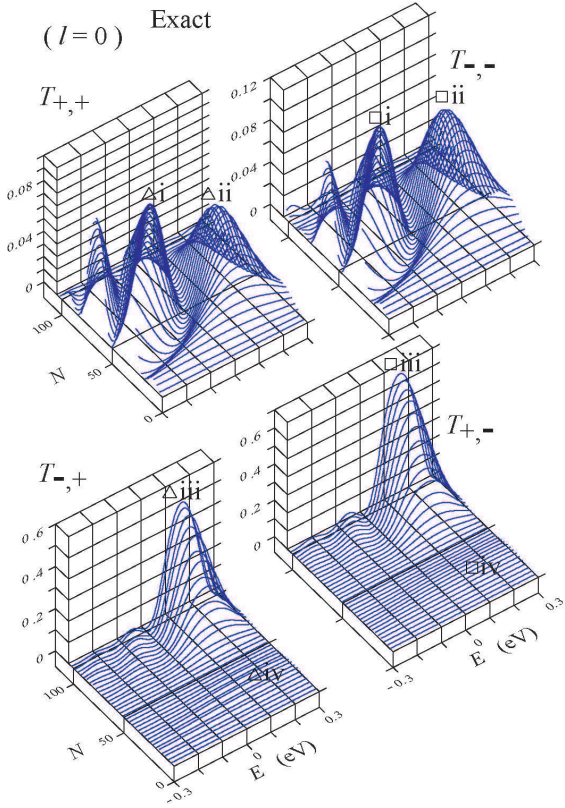} }}
\caption{
The same numerical data as in Fig. 9 except the residue $l=0$.
The local maximums are compared with those of Fig. 11 
in Table III.
The data $T$ of Fig. 11 are transformed by Eq. (\protect\ref{relation}) and shown in the rightmost column of Table IV.
}
\end{figure}

\begin{figure}
\epsfxsize=\columnwidth
\centerline{\hbox{ \epsffile{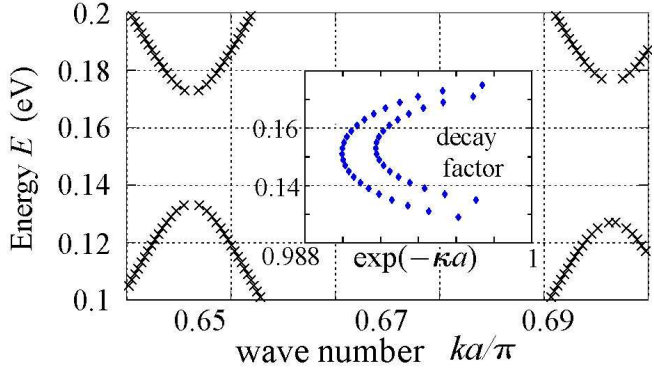} }}
\caption{ (main panel) The exact dispersion relation near the corner point.
(inset) The decay factors $e^{-\kappa a}$.
}
\end{figure}

\begin{figure}
\epsfxsize=\columnwidth
\centerline{\hbox{ \epsffile{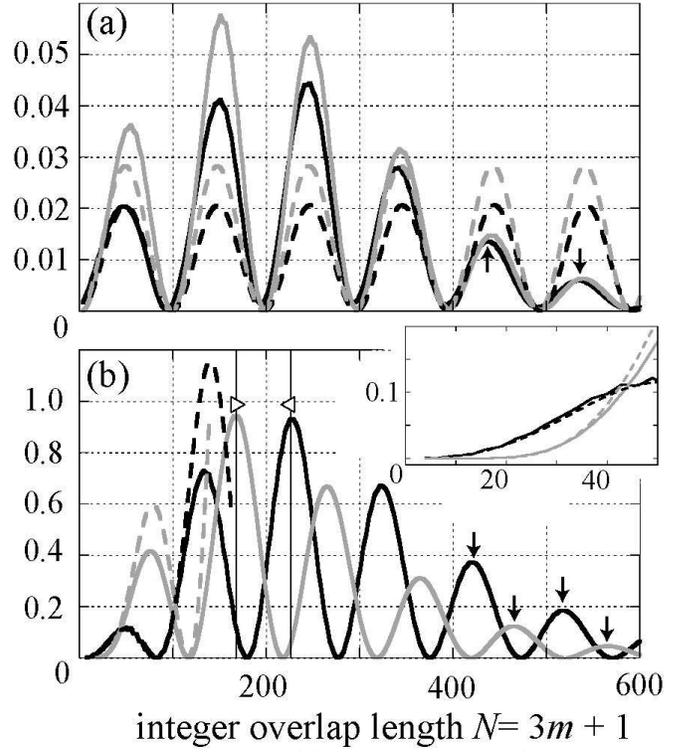} }}
\caption{
(a) $T_{\sigma,\sigma}$ and (b) $T_{-\sigma,\sigma}$
at $E=\varepsilon/2=0.15$ eV in case $l=1$.
The exact and approximate results are shown by
the solid and dashed lines, respectively.
Black and grey lines correspond to $\sigma= +$ and 
$\sigma= -$, respectively.
The neighboring peaks indicated by arrows
show the exponential decay of the gap state.
At the vertical line with $\triangleright$ ($\triangleleft$),
the VCF and the VCR manifest themselves as 
$T_{+,-} > 0.9$ and
$T_{-,+} < 0.1$ 
($T_{-,+} > 0.9$ and
$T_{+,-} < 0.1$). }
\end{figure}

\newpage

\begin{figure}
\epsfxsize=\columnwidth
\centerline{\hbox{ \epsffile{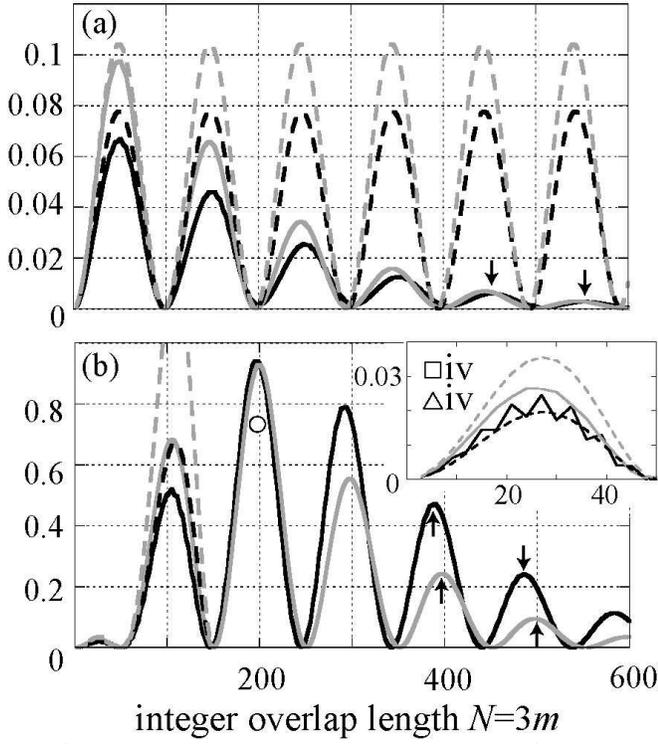} }}
\caption{The same as Fig. 15 except $l=0$.
At the peak with $\circ$,
the VCR manifest itself as 
$T_{+,-} \simeq T_{-,+} > 0.9$.}
\end{figure}

\begin{figure}
\epsfxsize=.9\columnwidth
\centerline{\hbox{ \epsffile{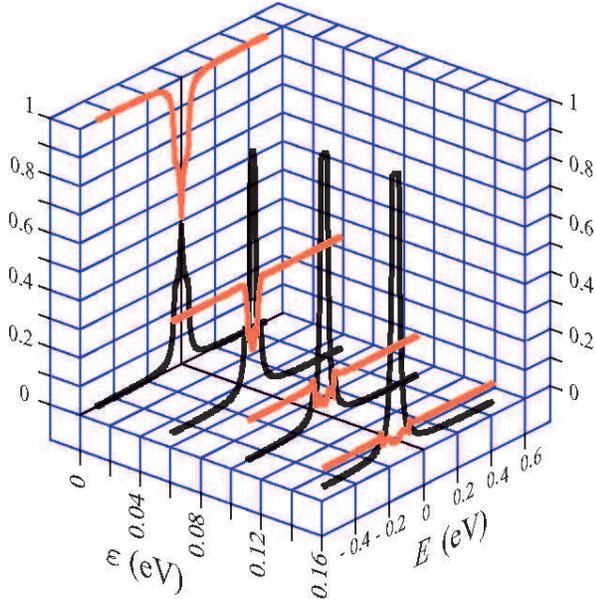} }}
\caption{ The product of the squared overlap integrals defined
by Eq. (\protect\ref{def-I}) where
$t_\bot=$ 0.04 eV and $\varepsilon=$ 0, 0.05, 0.1, 0.15 eV.
The black and red lines indicate
$I_{+,-}$ and $I_{+,+}$, respectively.
}
\label{Fig-I}
\end{figure}

\begin{figure}
\epsfxsize=.9\columnwidth
\centerline{\hbox{ \epsffile{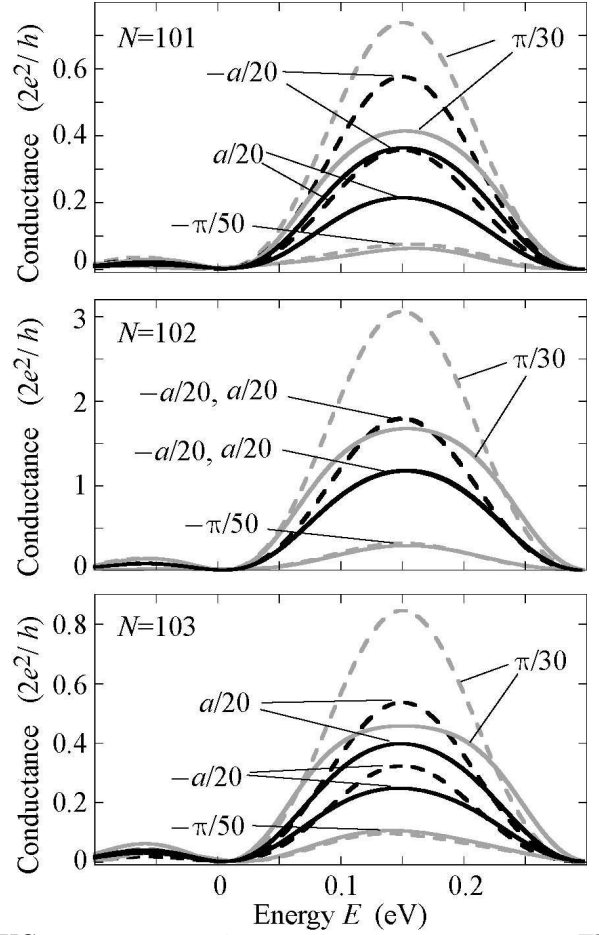} }}
\caption{The Landauer's formula conductance $G=(2e^2/h)\sum_{\sigma,\sigma'}T_{\sigma',\sigma}$ in case $(\theta_{\rm d},z_{\rm d})=(0,\pm a/20),(-\pi/50,0),(\pi/30,0)$, $N=$ 101, 102, 103.
The black and gray lines correspond to $\theta_{\rm d}=0$ and $z_{\rm d} =0$,
respectively.
The dashed lines correspond to Eqs. (\ref{perturb-T-offdiag}) and (\ref{SL2})
. The solid lines indicate the exact results.
}
\end{figure}


\begin{thebibliography}{10}
\bibitem{footnote}
The notations $A_{\sigma,\sigma}$ and $L$ in Ref. \protect\cite{tamura-2010} 
are represented by $A_{\sigma,\sigma}=4w_{\sigma,\sigma}^{(\sigma)}, L=(N-1)/2$
with the notations of the present paper.
Here $N=2L+1$ is limited to odd integers in Ref. \protect\cite{tamura-2010} 
while
we consider both odd $N$ and even $N$ in the present manuscript.
Since the rapid oscillation is suppressed in the average $\overline{T}(N)=(T(N-2)+T(N)+T(N+2))/3$, 
$L=(N-1)/2$ can be approximated by $L=N/2$ when $N \gg 1$.
\end{thebibliography}
\end{document}